\documentclass[12pt]{article}

\usepackage[a4paper,text={16.8cm,22.4cm}]{geometry}
\usepackage{amsmath,amsfonts,slashed,amssymb,tikz,bm,psfrag,graphicx,color,dsfont}
\usepackage{multicol}
\usepackage{float}

\RequirePackage[sort&compress,square,comma,numbers]{natbib}
\allowdisplaybreaks
\begin{document}

\begin{titlepage}

\begin{flushright}
\normalsize
March 18, 2018
\end{flushright}

\vspace{0.1cm}
\begin{center}
\Large\bf
Subleading-power corrections to the radiative leptonic $B \to \gamma \ell \nu$ decay in QCD
\end{center}

\vspace{0.5cm}
\begin{center}
{\bf Yu-Ming Wang$^{a,b}$,  Yue-Long Shen$^{c}$} \\
\vspace{0.7cm}
{\sl   ${}^a$ \, School of Physics, Nankai University, Weijin Road 94, 300071 Tianjin, China \\
 ${}^b$ \,  Fakult\"{a}t f\"{u}r Physik, Universit\"{a}t Wien, Boltzmanngasse 5, 1090 Vienna, Austria \\
 ${}^c$ \, College of Information Science and Engineering,
Ocean University of China, Songling Road 238, Qingdao, 266100 Shandong, P.R. China
}
\end{center}

\vspace{0.2cm}
\begin{abstract}

Applying the method of light-cone sum rules with photon distribution amplitudes,
we compute the subleading-power correction to the radiative leptonic $B \to \gamma \ell \nu$ decay,
at next-to-leading order in QCD for the twist-two contribution and at leading order in $\alpha_s$ for the higher-twist
contributions, induced by the  hadronic component of the collinear photon.
QCD factorization for the vacuum-to-photon correlation function with an interpolating current for
the $B$-meson is established explicitly at leading power in $\Lambda/m_b$ employing the evanescent operator approach.
Resummation of the parametrically large logarithms of $m_b^2/\mu^2$ entering the hard function of the leading-twist
factorization formula is achieved by solving the QCD evolution equation for the light-ray tensor operator at two loops.
The leading-twist hadronic photon effect turns out to preserve the symmetry relation between the two $B \to \gamma$
form factors due to the helicity conservation, however, the higher-twist hadronic photon corrections can yield symmetry-breaking
effect already at tree level in QCD. Using the conformal expansion of photon distribution amplitudes with
the non-perturbative parameters estimated from QCD sum rules, the twist-two hadronic photon contribution can
give rise to approximately 30\% correction to the leading-power ``direct photon" effect computed from
the perturbative QCD factorization approach. In contrast, the subleading-power corrections from the higher-twist two-particle
and three-particle photon distribution amplitudes are estimated to be of ${\cal O} (3 \sim 5\%)$ with the light-cone sum rule approach.
We further predict the partial branching fractions of $B \to \gamma \ell \nu $  with a photon-energy cut
$E_{\gamma} \geq E_{\rm cut}$, which are of interest for determining the inverse moment of the leading-twist $B$-meson
distribution amplitude thanks to the forthcoming high-luminosity Belle II experiment at KEK.

\end{abstract}

\vfil

\end{titlepage}

\section{Introduction}
\label{sect:Intro}

Exploring the subleading-power contributions to  exclusive $B$-meson decays in effective field
theories are of essential importance to understand  general properties of the
heavy quark expansion and its higher-order behaviours in QCD and to achieve precision determinations
of CKM matrix elements with a wealth of data accumulated at the $B$ factories
and at the LHC phenomenologically. In these respects, the radiative leptonic decay $B \to \gamma \ell \nu$ with an energetic
photon in the final state is widely believed to provide a clean probe of the strong interaction dynamics
of a heavy quark system and to put stringent constraints on the inverse moment of the leading-twist
$B$-meson distribution amplitude (DA). Factorization properties of $B \to \gamma \ell \nu$ have been investigated
extensively at leading power in $\Lambda/m_b$ with distinct QCD techniques \cite{Korchemsky:1999qb,DescotesGenon:2002mw}
and with the soft-collinear effective theory (SCET) \cite{Lunghi:2002ju,Bosch:2003fc,Beneke:2003pa}
which established the corresponding QCD factorization formula to all orders in perturbation theory.

Subleading-power corrections to the $B \to \gamma \ell \nu$ transition form factors were discussed in
QCD factorization at tree level \cite{Beneke:2011nf}, where the symmetry-preserving form factor $\xi(E_{\gamma})$ was introduced
to parameterize the non-local SCET matrix element without integrating out the hard-collinear scale.
Systematic studies on the higher-power terms of the radiative leptonic $B$-meson decay amplitude in the heavy quark expansion
are, however, still absent  in the framework of SCET beyond the leading-order in $\alpha_s$.
Applying the  dispersion relations and the parton-hadron duality, an alternative approach
without identifying manifest structures of the subleading-power effective operators
was proposed \cite{Braun:2012kp} to estimate the power suppressed soft contributions
at tree level and was further extended \cite{Wang:2016qii} to compute the  soft-overlap contribution at
next-to-leading-order (NLO) in QCD. Consequently, there will be a price to pay for the dispersion approach
taking into account the hadronic photon corrections and the end-point contributions (the so-called Feynman mechanism)
by implementing the non-perturbative modifications of the QCD spectral densities, as two additional non-perturbative
parameters (vector meson mass $m_{\rho}$ and  effective threshold parameter $s_{0}$) are introduced when
compared to the direct QCD calculation. It is then evident that evaluating the higher-power terms
in the expansion of $\Lambda/m_b$  individually  with direct QCD approaches is of particular interest to
deepen our understanding of perturbative QCD factorization for hard exclusive reactions.

The major objective of this paper is to perform QCD calculations of the subleading-power corrections
induced by the hadronic component  of the energetic photon at NLO in  the strong coupling constant.
QCD factorization formula for the two-particle hadronic photon correction to the $B \to \gamma \ell \nu$ amplitude
was demonstrated to be invalidated by the rapidity divergence in the convolution integral of the hard scattering
kernel with  the light-cone DAs  of the $B$-meson and of the photon \cite{Beneke:2003pa}.
Employing the technique of light-cone sum rules (LCSR) with the two-particle photon DAs, the power suppressed
``resolved photon" contribution was computed at twist-four accuracy and at leading-order (LO) in $\alpha_s$
\cite{Khodjamirian:1995uc,Ali:1995uy,Eilam:1995zv},
and was further updated \cite{Ball:2003fq} by including the NLO correction to the leading-twist hadronic photon DA contribution
and by calculating the higher-twist correction from the three-particle photon DAs at tree level.
However, QCD factorization for the vacuum-to-photon correlation function with an interpolating current for the
$B$-meson is not explicitly demonstrated with the operator-product-expansion (OPE) technique
at one loop in \cite{Ball:2003fq}, where the renormalization scheme dependence  of $\gamma_5$ for the QCD amplitude
in dimensional regularization was  not addressed in any detail.
It is therefore necessary to perform an independent calculation of the twist-two hadronic photon correction
to the $B \to \gamma \ell \nu$ form factors at NLO in $\alpha_s$ by compensating the above-mentioned gaps.
To this end, we will apply the standard perturbative matching procedure including the evanescent SCET operators
to establish QCD factorization formulae for the vacuum-to-$B$-meson correlation function
with the Dirac matrix $\gamma_5$ defined in naive dimensional
regularization (NDR) 
(see \cite{Bonneau:1990xu,Collins:1984xc} for an overview,
and \cite{Wang:2017ijn} for a discussion in the  context of the pion-photon transition form factor).

The presentation is organized as follows. We first summarize the theoretical status on  QCD calculations of
the $B \to \gamma \ell \nu$ form factors with different techniques based upon the heavy quark expansion
and discuss the origin of subleading-power corrections in section \ref{sect:Overview}.
To construct the sum rules for the leading-twist hadronic photon correction, we then establish QCD
factorization for the correlation function defined with an interpolating current for the $B$-meson
and with the weak transition current $ [\bar u \, \gamma_{\mu} \, (1-\gamma_5) \, b ]$ in section \ref{sect:leading-twist-photon},
where the master formula of the hard matching coefficient entering the factorization formula
at one loop will be derived with the  implementation of  the  infrared (IR) subtraction including the
evanescent SCET operator. With the aid of the evolution equation of the twist-two photon DA at two loops,
summation of the parametrically large logarithms of $m_b^2/\mu^2$ in the hard function will be further preformed
at next-to-leading-logarithmic (NLL) accuracy applying the  momentum-space renormalization group (RG) approach.
The NLL resummation improved LCSR for the twist-two hadronic correction to the $B \to \gamma$ form factors will be
also presented here, taking advantage of the dispersion relation technique and the parton-hadron duality ansatz.
The subleading-power corrections to the $B \to \gamma \ell \nu$ decay amplitude
from both the two-particle and three-particle higher-twist photon DAs displayed in \cite{Ball:2002ps} will be computed with the LCSR approach
at tree level in section \ref{sect:higher-twist-photon}, where a comparison of our results with
that obtained in  \cite{Eilam:1995zv,Ball:2003fq} will be also presented.
Phenomenological impacts of the various subleading-power corrections 
with the non-perturbative parameters of the photon DAs determined from QCD sum rules \cite{Balitsky:1989ry}
will be explored in section \ref{sect:numerical-analysis}, including the dependence of the
the partial branching fractions of $B \to \gamma \ell \nu$, with the phase-space cut of the
photon energy, on  the inverse moment $\lambda_B$.
A summary of our main observations and future perspectives  will be presented in section \ref{sect:Conclu}.
We further collect spectral representations of the convolution integrals entering the leading-twist factorization formulae
for the vacuum-to-photon correlation function at one-loop accuracy and the operator-level definitions
of the higher-twist photon DAs up to the twist-four in appendices \ref{app:spectral resp}
and \ref{app:Higher-twist photon DAs}, respectively.

\section{Theoretical overview of $B \to \gamma \ell \nu$ decay}
\label{sect:Overview}

The radiative leptonic $B \to \gamma \ell \nu$ decay amplitude is defined by the
QCD matrix element
\begin{eqnarray}
{\cal A}(B^{-} \to \gamma \, \ell \, \nu )
=\frac{G_F \, V_{ub}} {\sqrt{2}} \, \left \langle \gamma(p) \, \ell(p_{\ell}) \, \nu(p_{\nu}) \left |
\left [ \bar{\ell} \, \gamma_{\mu} \, (1- \gamma_5) \, \nu  \right ] \,\,
\left [ \bar u \, \gamma^{\mu} \, (1- \gamma_5) \, b \right ] \right | B^{-}(p_B) \right \rangle  \,.
\label{def: full decay amplitude}
\end{eqnarray}
Following \cite{Wang:2016qii} we will work in the  rest frame of the $B$-meson with momentum
$p_B = m_B \, v$ and introduce two light-cone vectors $n_{\mu}$ and $\bar{n}_{\mu}$ with the definitions
\begin{eqnarray}
p_{\mu}=\frac{n \cdot p}{2}\, \bar{n}_{\mu} \equiv E_{\gamma} \, \bar{n}_{\mu}\,,
\qquad v_{\mu} =\frac{ n_{\mu} + \bar{n}_{\mu}} {2} \,.
\end{eqnarray}
Expanding ${\cal A}(B^{-} \to \gamma \, \ell \, \nu )$ to the leading order in electromagnetic
interaction and employing the Ward identity due to the conservation of vector current leads to
\cite{Beneke:2011nf,Wang:2016qii}
\begin{eqnarray}
{\cal A}(B^{-} \to \gamma \, \ell \ \nu)
 \rightarrow  {G_F \, V_{ub} \over \sqrt{2}} \, \left ( i \, g_{em} \, \epsilon_{\nu}^{\ast}  \right ) \,
v \cdot p \, \bigg \{ - i  \, \epsilon_{\mu \nu \rho \sigma}
\, n^{\rho} \, v^{\sigma} \, F_V(n \cdot p)
+ g_{\mu \nu} \,  F_A(n \cdot p)  \bigg \} \,,
\label{B to photon form factor}
\end{eqnarray}
where the contribution due to photon radiation off the final-state lepton
has been taken into account by the redefinition of the axial form factor $F_A(n \cdot p)$.
It needs to be pointed out that the Lorentz indices $\mu$ and $\nu$ are transverse
relative to the four-vectors $v$ and $n$.

At leading power in $\Lambda/m_b$ the QCD factorization formula for the $B \to \gamma$
form factors can be readily derived  with the SCET technique \cite{Lunghi:2002ju,Bosch:2003fc}
\begin{eqnarray}
F_{V,\, \rm LP}(n \cdot p) = F_{A, \, \rm LP} (n \cdot p)= {Q_u \, m_B \over n \cdot p} \, \tilde{f}_B(\mu) \,
C_{\perp}(n \cdot p, \mu) \, \int_0^{\infty} \, d \omega \, {\phi_B^{+}(\omega, \mu) \over \omega} \,
J_{\perp}(n \cdot p,\omega,  \mu) \,.
\label{leading-power factorization formula}
\end{eqnarray}
The hard function $C_{\perp}$ arises from matching the QCD weak current
$\bar u \, \gamma_{\mu \, \perp} \, (1-\gamma_5) \, b$ onto the corresponding
SCET current and the one-loop expression is given by \cite{Bauer:2000yr,Beneke:2004rc}
\begin{eqnarray}
C_{\perp}&=& 1- \frac{\alpha_s \, C_F}{4 \, \pi}
\bigg [ 2 \, \ln^2 {\mu \over n \cdot p} + 5 \, \ln {\mu \over m_b}
-2 \, {\rm Li}_2 \left ( 1-{1 \over r} \right )  - \ln^2  r  \nonumber \\
&&  + \,  {3 r -2 \over 1 -r}  \, \ln r + {\pi^2 \over 12} + 6  \bigg ]  \,,
\end{eqnarray}
with $r=n \cdot p /m_b$.
The hard-collinear function $J_{\perp}$ entering the SCET factorization formula
(\ref{leading-power factorization formula}) reads \cite{Lunghi:2002ju,Bosch:2003fc,Wang:2016qii}
\begin{eqnarray}
J_{\perp} =  1 + {\alpha_s \, C_F \over 4 \, \pi} \,
\left [ \ln^2 { \mu^2 \over n \cdot p \,  (\omega - \bar n \cdot p)}  - {\pi^2 \over 6} - 1 \right ]
+ {\cal O}(\alpha_s^2) \,.
\end{eqnarray}
Setting $\mu$ as a hard-collinear scale of order $\sqrt{\Lambda\, m_b}$ and performing the NLL resummation
of the parametrically large logarithms in the hard function yields
\begin{eqnarray}
F_{V,\, \rm LP}(n \cdot p) &=& F_{A, \, \rm LP} (n \cdot p) \nonumber \\
&=& {Q_u \, m_B \over n \cdot p \,\, \lambda_B(\mu)} \,
\left [  U_2(n \cdot p, \mu_{h2}, \mu) \,\tilde{f}_B(\mu_{h2}) \right ] \,
\left [ U_1(n \cdot p, \mu_{h1}, \mu) \, C_{\perp}(n \cdot p, \mu_{h1})   \right ]\, \nonumber \\
&& \times \bigg \{  1 + {\alpha_s(\mu) \, C_F \over 4 \,  \pi} \,
\bigg [\sigma_2(\mu) + 2 \, \ln {\mu^2 \over n \cdot p \, \mu_0} \, \sigma_1(\mu)
+ \ln^2 {\mu^2 \over n \cdot p \, \mu_0} - {\pi^2 \over 6} -1 \bigg ] \bigg \}  \,,
\label{resummation improved leading-power factorization formula}
\end{eqnarray}
where the convolution integral of $\omega$ has been expressed as  moments of the $B$-meson DA
defined in \cite{Beneke:2011nf} and the manifest  expressions of  the evolution functions
$U_1$ and $U_2$ can be found in \cite{Wang:2016qii}.

The subleading-power corrections from  photon radiation off the heavy quark and
from  higher-twist $B$-meson DAs were addressed  \cite{Beneke:2011nf} by computing the two
diagrams for the tree $b \, \bar u \to \gamma \, W^{\ast}$ amplitude in QCD.
Since the factorization property of the non-local subleading-power correction from photon radiation off the light quark
has not been explored yet, we will only focus on the local subleading-power contribution
to the $B \to \gamma \ell \nu$ amplitude at tree level
\begin{eqnarray}
F_{V,\, \rm NLP}^{\rm LC}(n \cdot p)=-F_{A,\, \rm NLP}^{\rm LC}(n \cdot p)
= \frac{Q_u \, f_B \, m_B}{(n \cdot p)^2}
+ \frac{Q_b \, f_B \, m_B}{n \cdot p \, m_b}  \,.
\label{local subleading power corrections}
\end{eqnarray}
As discussed in \cite{Beneke:2003pa} the subleading-power contribution can be further generated by the
effective matrix element $\langle \gamma(p) | {\cal O} | B^{-}(p_B) \rangle $ with the SCET operator
${\cal O} \supset [\bar q_s \, h_v]_s \,\, [\bar \xi \, \xi]_c$ containing no photon field,
due to the unsuppressed interactions of photos with any numbers of collinear quark
and gluon  fields. The collinear  matrix element $\langle \gamma(p) | [\bar \xi \, \xi]_c | 0 \rangle $
defines the photon DAs on the light cone, making the photon behave in analogy to  an energetic vector meson.
Consequently, these terms are also referred to as the ``hadronic (resolved) photon" contributions in different contexts.
QCD calculations of such power suppressed corrections to the $B \to \gamma \ell \nu$ decay form factors will be
carried out, to the twist-four photon DAs accuracy,  with the LCSR approach in the following.

\section{Leading-twist hadronic photon correction in QCD}
\label{sect:leading-twist-photon}

To obtain the sum rules for the form factors $F_{V}(n \cdot p)$ and $F_{A}(n \cdot p)$,
we construct the vacuum-to-photon correlation function with an interpolating current
for the $B$-meson
\begin{eqnarray}
\Pi_{\mu}(p, q) = \int d^4 x \, e^{i \, q \cdot x} \,
\langle \gamma(p) | T \{ \bar u(x) \, \gamma_{\mu \, \perp} \, (1- \gamma_5) \, b(x), \,
\bar b(0) \, \gamma_5 \, u(0) \} | 0 \rangle \,,
\label{def: correlation function}
\end{eqnarray}
where $q=p_{\ell}+p_{\nu}$ refers to the four-momentum of the lepton-neutrino pair.
QCD factorization for the correlation function (\ref{def: correlation function}) can be demonstrated
with the technique of OPE at $(p+q)^2 \ll m_b^2$ and $q^2 \ll m_b^2$.
For definiteness, we will employ the following power counting scheme
\begin{eqnarray}
n \cdot p \sim {\cal O}(m_b), \qquad | n \cdot (p+q) - m_b | \sim {\cal O}(\Lambda)\,.
\label{power counting scheme}
\end{eqnarray}
The primary task of this section is to compute the perturbative matching coefficient entering
the leading-twist factorization formula for  (\ref{def: correlation function}) at NLO,
with the evanescent operator approach.

\subsection{The twist-two hadronic photon correction at tree level}

\begin{figure}
\begin{center}
\includegraphics[width=0.50 \columnwidth]{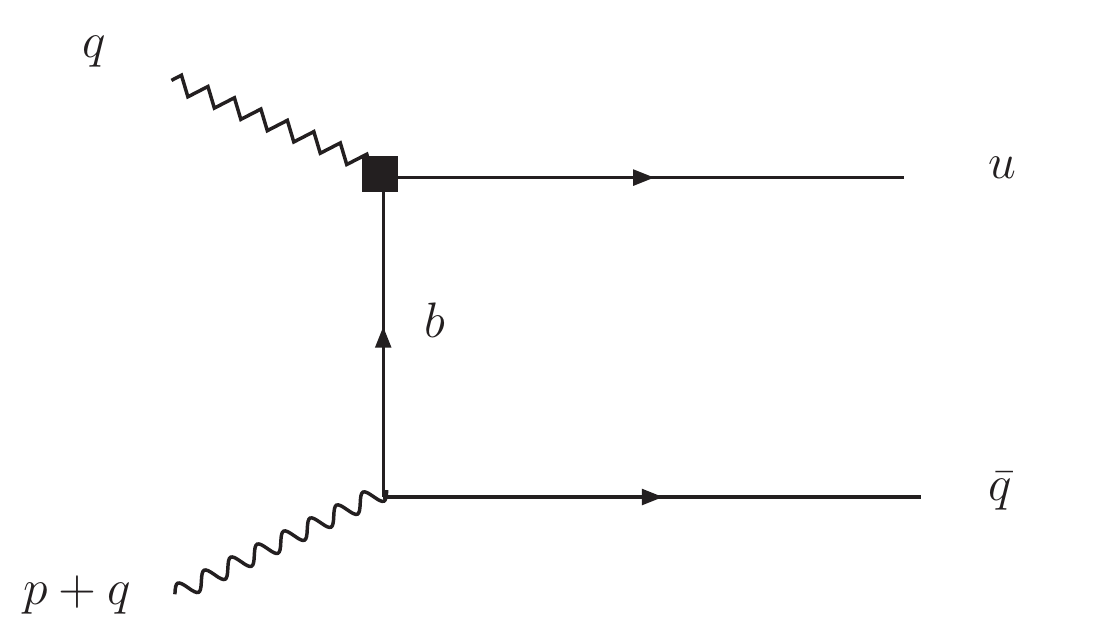} \\
\vspace*{0.1cm}
\caption{Diagrammatical representation of the leading-order (LO) contribution to
the QCD amplitude $F_{\mu}(p, q)$ defined in (\ref{QCD amplitude at twist-2}). }
\label{fig:twist-2 correlator at LO}
\end{center}
\end{figure}

QCD factorization for the twist-two contribution to the correlation function  (\ref{def: correlation function})
can be justified by investigating the four-point QCD amplitude
\begin{eqnarray}
F_{\mu}(p, q) = \int d^4 x \, e^{i \, q \cdot x} \,
\langle q(z \, p) \, \bar q (\bar z \, p) | T \{ \bar u(x) \, \gamma_{\mu \, \perp} \, (1- \gamma_5) \, b(x), \,
\bar b(0) \, \gamma_5 \, u(0) \} | 0 \rangle \,,
\label{QCD amplitude at twist-2}
\end{eqnarray}
where $z$ indicates the momentum fraction carried by the collinear quark and $\bar z \equiv 1-z$.
Evaluating the tree diagram displayed in figure \ref{fig:twist-2 correlator at LO} leads to
\begin{eqnarray}
F_{\mu}^{(0)}(p, q) &=& {i \over 2} \,  \frac{\bar n \cdot q}{z (p+q)^2 + \bar z \, q^2 - m_b^2 + i 0} \,
\bar u(z \, p) \, \gamma_{\mu \perp} \, \not \! n \, (1 + \gamma_5)  \, v(\bar z \, p) \nonumber \\
&=& {i \over 2} \,  \frac{\bar n \cdot q}{z^{\prime} (p+q)^2 + \bar z^{\prime}  \, q^2 - m_b^2 + i 0}
\ast \langle O_{A, \, \mu}(z, z^{\prime}) \rangle^{(0)}\,,
\end{eqnarray}
where the convolution integral of $z^{\prime}$ is represented by an asterisk.
$\langle O_{A, \, \mu}(z, z^{\prime}) \rangle^{(0)}$ indicates the partonic matrix element
of the SCET operator $O_{A, \, \mu}$ at tree level
\begin{eqnarray}
\langle O_{A, \, \mu}(z, z^{\prime}) \rangle=
\langle q(z \, p) \, \bar q (\bar z \, p) |  O_{A, \, \mu}(z^{\prime})| 0 \rangle
= \bar \xi(z \, p) \, \gamma_{\mu \perp} \, \not \! n \, (1 + \gamma_5)\, \xi(z \, p) \,
\delta(z-z^{\prime}) + {\cal O}(\alpha_s), \hspace{0.3 cm}
\end{eqnarray}
where the general definition of the collinear operator in moment space reads
\begin{eqnarray}
O_{j, \, \mu}(z^{\prime}) &=& \frac{n \cdot p}{2 \, \pi} \, \int d \tau \,
e^{-i \,z^{\prime} \, \tau \, n \cdot p } \,
\bar \xi(\tau \, n) \, W_c(\tau \, n,0) \,
\Gamma_j \,  \xi(0) \,, \nonumber \\
\Gamma_j &=&  \gamma_{\mu \perp} \, \not \! n \, (1 + \gamma_5) \,.
\end{eqnarray}
The collinear Wilson line with the convention of the covariant derivative
$D_{\mu} \equiv \partial_{\mu} - i \, g_s \, A_{\mu}$ is defined as
\begin{eqnarray}
W_c(\tau  n,0) = {\rm P} \, \left \{ {\rm  Exp} \left [   i \, g_s \,
\int_{0}^{\tau} \, d \lambda \,   n  \cdot A_{c}(\lambda \, n) \right ]  \right \} \,.
\end{eqnarray}

To establish the hard-collinear factorization for the QCD amplitude (\ref{QCD amplitude at twist-2}),
we further decompose the SCET operator $O_{\mu}$ into the light-ray operators defining the photon DAs
displayed in \cite{Ball:2002ps}
\begin{eqnarray}
O_{A, \, \mu} &=& O_{1, \, \mu} + O_{2, \, \mu} + O_{E, \, \mu}  \,,
\end{eqnarray}
with
\begin{eqnarray}
\Gamma_{1} &=&   \gamma_{\mu \perp} \, \not \! n  \,, \qquad
\Gamma_{2}= {n^{\nu} \over 2} \, \epsilon_{\mu \nu \alpha \beta} \, \sigma^{\alpha \beta} \,, \qquad
\Gamma_{E}= \gamma_{\mu \perp} \, \not \! n  \, \gamma_5 -
{n^{\nu} \over 2} \, \epsilon_{\mu \nu \alpha \beta} \, \sigma^{\alpha \beta} \,.
\end{eqnarray}
It is evident that $O_{E, \, \mu}$ is an evanescent operator vanishing in four-dimensional space.
Expanding the operator matching equation including the  evanescent operator
\begin{eqnarray}
F_{\mu}(p,q)= \sum_i \, C_i(z^{\prime}, (p+q)^2, q^2) \ast \langle O_{1, \, \mu}(z, z^{\prime})  \rangle \,,
\label{matching condition}
\end{eqnarray}
to the LO in the strong coupling constant, gives rise to
\begin{eqnarray}
C_1^{(0)} = C_2^{(0)}  = C_E^{(0)}
= {i \over 2} \,  \frac{\bar n \cdot q}{z^{\prime} (p+q)^2 + \bar z^{\prime}  \, q^2 - m_b^2 + i 0} \,.
\end{eqnarray}
Taking advantage of the definition of the leading-twist photon DA \cite{Ball:2002ps}
\begin{eqnarray}
&& \langle \gamma(p) |\bar \xi(x) \, W_c(x, 0) \,\, \sigma_{\alpha \beta} \,\, \xi(0)| 0 \rangle  \nonumber \\
&& = i \, g_{\rm em} \, Q_q \, \chi(\mu) \, \langle \bar q q \rangle(\mu) \,
(p_{\beta} \, \epsilon_{\alpha}^{\ast} - p_{\alpha} \, \epsilon_{\beta}^{\ast}) \,
\int_0^1 \, d z \, e^{i \, z \, p \cdot x} \, \phi_{\gamma}(z, \mu) + {\cal O}(x^2) \,.
\end{eqnarray}
we can readily derive the tree-level factorization formula for the correlation function
(\ref{def: correlation function})
\begin{eqnarray}
\Pi_{\mu}(p, q)&=&  {i  \over 2} \, g_{\rm em} \, Q_u \,
\chi(\mu) \, \langle \bar q q \rangle(\mu) \,\epsilon^{\ast \, \alpha}(p)
\left [ g_{\mu \alpha}^{\perp} - i \, \epsilon_{\mu \alpha \nu \beta} \,
n^{\nu} \, v^{\beta} \right ] \, \nonumber \\
&& \times \int_0^1 \, d z \, \phi_{\gamma}(z \,, \mu) \,\,
\frac{n \cdot p \,\, \bar n \cdot q}{z (p+q)^2 + \bar z  \, q^2 - m_b^2 + i 0} +{\cal O}(\alpha_s)  \,.
\label{tree level factorization formula at LT}
\end{eqnarray}

Employing the definition of the $B$-meson decay constant in QCD
\begin{eqnarray}
\langle B^{-}(p_B) | \bar b \, \gamma_5 \, u | 0\rangle = - i \,
{f_B \, m_B^2 \over m_b+m_u} \,,
\end{eqnarray}
we can derive the hadronic dispersion relation of (\ref{def: correlation function}) as follows
\begin{eqnarray}
\Pi_{\mu}(p, q) &=& {i  \over 2} \, g_{\rm em} \, {f_B\, m_B^2 \over m_b+m_u} \, \epsilon^{\ast \, \alpha}(p) \,
\left [ g_{\mu \alpha}^{\perp} \, F_{A, \, \rm photon}^{\rm 2PLT}(n \cdot p)
- i \, \epsilon_{\mu \alpha \nu \beta} \,n^{\nu} \, v^{\beta} \, \, F_{V, \, \rm photon}^{\rm 2PLT}(n \cdot p) \right ] \,
\nonumber \\
&& \times {n \cdot p  \over (p+q)^2 - m_B^2 + i 0}
\, + \int_{s_0}^{\infty} \, ds \, {\rho^{h}_{\mu}(s, q^2) \over s - (p+q)^2 - i 0} \,,
\label{hadronic dispersion relation}
\end{eqnarray}
where $s_0$ is the effective threshold of the $B$-meson channel. The tree-level LCSR for the $B \to \gamma \ell \nu$
form factors can be obtained by matching the factorization formula (\ref{tree level factorization formula at LT})
and (\ref{hadronic dispersion relation}) with the aid of the parton-hadron duality approximation
and the Borel transformation
\begin{eqnarray}
{f_B\, m_B \over m_b+m_u} \, F_{V, \, \rm photon}^{\rm 2PLT}(n \cdot p)
&=&{f_B\, m_B \over m_b+m_u} \, F_{A, \, \rm photon}^{\rm 2PLT}(n \cdot p) \nonumber \\
&=& Q_u \, \chi(\mu) \, \langle \bar q q \rangle(\mu) \,
\int_{z_0}^1 \, {d z \over z} \, {\rm exp}
\left [ -{m_b^2 - \bar z \, q^2 \over z \, M^2}+ {m_B^2 \over M^2} \right ] \, \phi_{\gamma}(z, \mu) \nonumber \\
&& + \, {\cal O}(\alpha_s)\,,
\label{tree-level sum rule at twist-2}
\end{eqnarray}
with $z_0=(m_b^2-q^2)/(s-q^2)$. With the power counting scheme for the threshold parameter and the Borel mass
entering the sum rules (\ref{tree-level sum rule at twist-2})
\begin{eqnarray}
\left ( s_0-m_b^2 \right ) \sim M^2 \sim {\cal O}(m_b \, \Lambda) \,, \qquad  \bar z_0 \sim {\Lambda / m_b} \,,
\label{power counting for the threshold}
\end{eqnarray}
the heavy-quark scaling of the hadronic photon correction at leading twist can be established
\begin{eqnarray}
F_{V, \, \rm photon}^{\rm 2PLT} \sim  F_{A, \, \rm photon}^{\rm 2PLT}
\sim {\cal O} \left ({\Lambda \over m_b} \right )^{3/2} \,,
\label{power counting of the twist-2 hadronic photon contribution}
\end{eqnarray}
which is indeed suppressed by a factor of $\Lambda/m_b$ compared with the direct photon contribution
\begin{eqnarray}
\left \langle \gamma(p) \left | \bar q_s \, \not \! \! A_{\perp (\gamma)} \,
{1 \over i \, \bar n \cdot \overleftarrow{D}_s} {\not \!  {\bar n} \over 2} \,
\gamma_{\mu} \, (1-\gamma_5)  \, h_v \right | B^{-}(p_B) \right \rangle
\sim {\cal O} \left ({\Lambda \over m_b} \right )^{1/2}  \,.
\label{power counting of the LP contribution}
\end{eqnarray}

\subsection{The twist-two hadronic photon correction at one loop}

\begin{figure}
\begin{center}
\includegraphics[width=1.0 \columnwidth]{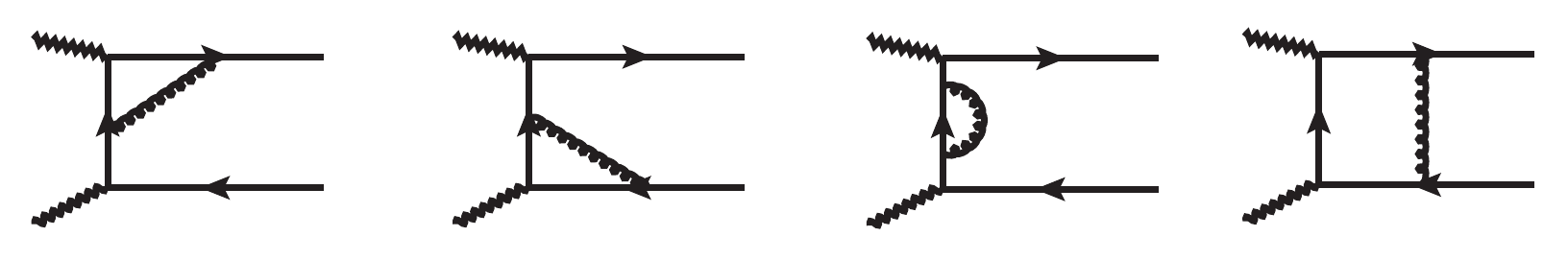} \\
(a) \hspace{3.5 cm} (b) \hspace{3.5 cm} (c) \hspace{3.5 cm} (d)\\
\vspace*{0.1cm}
\caption{Diagrammatical representation of the NLO contribution to
the QCD amplitude $F_{\mu}(p, q)$ defined in (\ref{QCD amplitude at twist-2}). }
\label{fig:twist-2 correlator at NLO}
\end{center}
\end{figure}

In this subsection we will proceed to derive the NLO sum rules for the twist-two hadronic
photon correction to the $B \to \gamma$ form factors and to perform  resummation of
the large logarithms of $m_b^2/\mu^2$ in the hard function at NLL accuracy.
To this end, we will  need to demonstrate QCD factorization for the vacuum-to-photon correlation
function (\ref{def: correlation function}) at one loop,  applying the technique of the light-cone OPE.
For the sake of determining the NLO matching coefficients entering the factorization formulae of $\Pi_{\mu}(p, q)$,
we will first evaluate the one-loop diagrams for the QCD matrix element $F_{\mu}(p, q)$ displayed
in figure \ref{fig:twist-2 correlator at NLO}.

The one-loop QCD correction to the weak vertex diagram shown in figure \ref{fig:twist-2 correlator at NLO}(a)
can be readily computed as
\begin{eqnarray}
F_{\mu, \, \rm weak}^{(1)} &=& \frac{g_s^2 \, C_F}{z (p+q)^2 + \bar z  \, q^2 - m_b^2 + i 0}  \,
\int {d^D l \over (2 \, \pi)^D} \, \frac{1}{[(z \, p + l)^2 + i 0] [(z \, p + q + l)^2 - m_b^2 + i 0] [l^2 + i 0]} \nonumber \\
&& \bar u(z \, p) \, \gamma_{\nu} \, (z \not \! p + \not  l) \, \gamma_{\mu \perp} \, (1-\gamma_5)\,
(z \not \! p + \not \! q +  \not  l +m_b) \, \gamma^{\nu} \, (z \not \! p +  \not  l +m_b) \, \gamma_5 \, v(\bar z \, p)\,,
\end{eqnarray}
where the external partons are already taken to be on the mass-shell due to the insensitivity of the hard matching coefficients
on the IR physics. With the power counting scheme specified in  (\ref{power counting scheme}),
one can identify the leading-power contributions of the scalar integral
\begin{eqnarray}
I_1 = \int {d^D l \over (2 \, \pi)^D} \, \frac{1}{[(z \, p + l)^2 + i 0] [(z \, p + q + l)^2 - m_b^2 + i 0] [l^2 + i 0]} \,,
\end{eqnarray}
from the hard and collinear regions as expected. Applying the method of regions \cite{Beneke:1997zp},
the collinear contribution of $I_1$ vanishes in dimensional regularization due to the resulting scaleless integral,
and will be cancelled by the corresponding IR subtraction term independent of the regularization scheme.
Reducing the Dirac algebra of $F_{\mu, \, \rm weak}^{(1)}$ with the NDR scheme of the Dirac matrix $\gamma_5$
and preforming the loop-momentum integration leads to
\begin{eqnarray}
F_{\mu, \, \rm weak}^{(1), \, h} &=& {\alpha_s \, C_F \over 4 \, \pi }  \,
\bigg \{  \left [ { 2 (1-r_2) \over r_2 -r_1 } \, \ln {1 - r_1 \over 1-r_2}  - 1 \right ] \,
\bigg  [ {1 \over \epsilon} + \ln {\mu^2 \over m_b^2} -{\ln [(1-r_1) (1-r_2)]  \over 2}
 - {r_1 - 3\, r_2 \over 4 \, (1-r_2)} + 2 \bigg ]  \nonumber \\
&& + \, {1 \over r_1 - r_2} \, \bigg [ 2 \, (1- r_2) \, {\rm Li_2}\left ( 1- {1 - r_1 \over 1-r_2}  \right )
- {2 \, [r_1 (r_1-2) + r_2] \over r_1} \, \ln (1-r_1)  \nonumber \\
&& + \, {2 \, [r_1 (r_2-2) + r_2] \over r_2} \, \ln (1-r_2)  + {r_2 \over 2} \bigg ]
- {r_1 -3 \, r_2 \over 4 \, (1- r_2)}  - 3 \bigg \}  \,
F_{\mu}^{(0)} \,,
\end{eqnarray}
where $r_1=(z \, p + q)^2/m_b^2$ and $r_2=q^2/m_b^2$.


Along the same vein, the one-loop QCD correction to the $B$-meson vertex diagram
displayed in figure \ref{fig:twist-2 correlator at NLO}(b) can be written as
\begin{eqnarray}
F_{\mu, \, B}^{(1)} &=& - \frac{g_s^2 \, C_F}{z (p+q)^2 + \bar z  \, q^2 - m_b^2 + i 0}  \,
\int {d^D l \over (2 \, \pi)^D} \, \frac{1}{[(\bar z \, p - l)^2 + i 0] [(z \, p + q + l)^2 - m_b^2 + i 0] [l^2 + i 0]} \nonumber \\
&& \bar u(z \, p) \, \gamma_{\mu \perp} \, (1-\gamma_5)\,
(z \not \! p + \not  \! q  +m_b) \, \gamma_{\nu} \, (z \not \! p + \not \! q +  \not  l +m_b) \, \gamma_5 \,
(\bar z \,  \not \! p -  \not  l ) \, \gamma^{\nu}  \, v(\bar z \, p)\,,
\end{eqnarray}
which again depends on the precise prescription of $\gamma_5$ in the complex
$D$-dimensional space.
It is straightforward to verify that the leading-power contributions to the $B$-meson
vertex diagram also arise from the hard and collinear regions.
Evaluating the hard contribution to  $F_{\mu, \, B}^{(1)}$ with  the method of regions
in the NDR scheme of $\gamma_5$ yields
\begin{eqnarray}
F_{\mu, \, B}^{(1), \, h} &=& - {\alpha_s \, C_F \over 4 \, \pi }  \,
\bigg \{ 2 \, \left [ {1-r_3 \over r_1-r_3} \, \ln {1 - r_1 \over 1-r_3} - 1 \right ] \,
\left [ {1 \over \epsilon} + \ln {\mu^2 \over m_b^2}
-{\ln [(1-r_1) (1-r_3)]  \over 2}  + {3 \, r_1 - r_3 \over 2 \, (1-r_3)} \right ] \nonumber \\
&& - {2 \over r_1 -r_3} \, \left [ (1-r_3) \, {\rm Li}_2 \left (1- {1- r_1 \over 1-r_3} \right )
+ (3 \, r_1 - r_3 -1) \, \left ( {\ln (1-r_1) \over r_1} - {\ln (1-r_3) \over r_3} \right ) \right ]  \nonumber \\
&& - {1- 3 \, r_1 \over 1-r_3}  + 3 \bigg \}  \,
F_{\mu}^{(0)} \,,
\end{eqnarray}
with $r_3=(p+q)^2/m_b^2$.


The self-energy correction to the intermediate bottom-quark propagator displayed
in figure \ref{fig:twist-2 correlator at NLO}(c)  can be  computed as
\begin{eqnarray}
F_{\mu, \, wfc}^{(1)} &=& - {\alpha_s \, C_F \over 4 \, \pi} \,
\left \{ {7 - r_1 \over 1- r_1} \, \left [ {1 \over \epsilon} + \ln {\mu^2 \over m_b^2}
- \ln (1-r_1) + {1 \over 2} \right ]\right \}   \,
F_{\mu}^{(0)} \,.
\end{eqnarray}
Furthermore, the wave function renormalization of the external quarks will be cancelled precisely
by the corresponding collinear subtraction term and hence will not contribute to the perturbative matching coefficients.


Now we turn to compute the one-loop correction to the box diagram displayed in figure
\ref{fig:twist-2 correlator at NLO}(d)
\begin{eqnarray}
F_{\mu, \, box}^{(1)} &=& - g_s^2 \, C_F \,
\int {d^D l \over (2 \, \pi)^D} \, \frac{1}{[(z \, p + l)^2 + i 0] [(z \, p + q + l)^2 - m_b^2 + i 0]
[(\bar u \, p - l)^2  + i 0] [l^2 + i 0]} \nonumber \\
&& \bar u(z \, p) \,  \gamma_{\nu} \, (z \not \! p + \not  l) \, \gamma_{\mu \perp} \, (1-\gamma_5)\,
(z \not \! p + \not  \! q  + \not  l + m_b)  \, \gamma_5 \, (\bar z \not \! p - \not  l) \,
 \gamma^{\nu}  \, v(\bar z \, p) \nonumber \\
&=&  - i \, g_s^2 \, C_F \,  F_{\mu}^{(0)}
\int {d^D l \over (2 \, \pi)^D} \, \frac{m_b^2 \, (r_1 - 1) }{[(z \, p + l)^2 + i 0] [(z \, p + q + l)^2 - m_b^2 + i 0]
[(\bar u \, p - l)^2  + i 0] [l^2 + i 0]} \nonumber  \\
&&  \times \, (D-4) \,  \left \{  - {D-4 \over D-2} \,\, l_{\perp}^2
+ {\bar n \cdot l \over \bar n \cdot q} \, \left [ \bar n \cdot l \,\, n \cdot (u \, p + q) + l^2 \right ] \right \} \,,
\end{eqnarray}
where the reduction of the Dirac algebra is achieved with the NDR scheme of $\gamma_5$ in the second step
and $l_{\perp}^2 \equiv g_{\mu \nu}^{\perp} \, l^{\mu} \, l^{\nu}$.
Performing the loop-momentum  integration we find that the one-loop box diagram only contributes
at ${\cal O}(\epsilon)$, vanishing in four dimensional space.
Such observation is in analogy to  the hard-collinear factorization for the hadronic photon correction
to the pion-photon form factor at leading-twist accuracy \cite{Wang:2017ijn}.

Adding  up different pieces together, we obtain the one-loop QCD correction to the four-point
QCD matrix element as follows
\begin{eqnarray}
F_{\mu}^{(1)}(p, q)
&=& T_{A, \, \rm hard}^{(1)}(z^{\prime}, (p+q)^2, q^2)
\ast \langle O_{A, \, \mu}(z, z^{\prime}) \rangle^{(0)} + ... \nonumber  \\
&=& \sum_{i=1, 2, E} \, T_{i, \, \rm hard}^{(1)}(z^{\prime}, (p+q)^2, q^2) \,
\ast \langle O_{i, \, \mu}(z, z^{\prime}) \rangle^{(0)} + ... \,,
\end{eqnarray}
where the explicit expression of the NLO hard amplitude is given by
\begin{eqnarray}
T_{i, \, \rm hard}^{(1)}\big |_{\rm NDR} &=& { \alpha_s \, C_F \over 4 \, \pi} \, \bigg \{ (-2) \,
\bigg [ {1 -r_2 \over r_1-r_2} \, \ln {1-r_1 \over 1-r_2}
+ {1 -r_3 \over r_1-r_3}  \, \ln {1-r_1 \over 1-r_3} + {3 \over 1-r_1}\bigg ]
\left ({1 \over \epsilon} + \ln {\mu^2 \over m_b^2} \right ) \nonumber \\
&& + {2 \, (1-r_2) \over r_1-r_2} \, {\rm Li}_2 \left (1-{1- r_1 \over 1-r_2} \right )
+ {2 \, (1-r_3) \over r_1-r_3} \, {\rm Li}_2 \left (1-{1- r_1 \over 1-r_3} \right ) \nonumber \\
&& + \left ( {1-r_2 \over r_1 -r_2} + {1-r_3 \over r_1 -r_3 }\right ) \, \ln^2 (1-r_1)
- {1-r_2 \over r_1-r_2} \, \ln^2 (1-r_2) - {1-r_3 \over r_1-r_3} \, \ln^2 (1-r_3) \,\nonumber \\
&& + \left [ {2 \over r_1 (r_3-r_1)} + { 2 (r_3-2) \over r_3-r_1}
+ {6 \over 1- r_1} - {2 - r_2 \over r_1 -r_2}  + {4 \over r_1} - 4 \right ]  \, \ln (1-r_1)\nonumber \\
&& + \left ( {2 - r_2 \over r_1 -r_2} - {4 \over r_2} + 2\right ) \, \ln (1-r_2)
+ \left [ {2 \over r_3 (r_1 -r_3)} + {2 \, (r_1-2) \over r_1-r_3} - {6 \over r_3} \right ]  \,
 \ln (1-r_3) \nonumber \\
&& + {r_2 \over 2 \, (r_1 - r_2)} - {3 \over 1 -r_1} - {15 \over 2} \bigg \} \, C_{i, \rm hard}^{(0)}\,,
\label{NLO hard contribution}
\end{eqnarray}
where the parameter $z$ in the definition of $r_1$ should be apparently understood as $z^{\prime}$.

We are now in a position to derive the master formulae for the hard functions
$C_{1, 2}(z^{\prime}, (p+q)^2, q^2)$ by implementing the ultraviolet (UV) renormalization
and  the IR subtraction. Expanding the operator matching condition (\ref{matching condition})
at ${\cal O}(\alpha_s)$ gives rise to
\begin{eqnarray}
&& \sum_i \, T_i^{(1)}(z^{\prime}, (p+q)^2, q^2) \ast \langle O_{i, \, \mu}(z, z^{\prime}) \rangle^{(0)} \nonumber \\
&& = \sum_i \, \left [ C_i^{(1)}(z^{\prime}, (p+q)^2, q^2) \ast \langle O_{i, \, \mu}(z, z^{\prime}) \rangle^{(0)}
+  C_i^{(0)}(z^{\prime}, (p+q)^2, q^2) \ast \langle O_{i, \, \mu}(z, z^{\prime}) \rangle^{(1)} \right ]. \,\,
\label{matching relation at NLO}
\end{eqnarray}
The UV renormalized one-loop SCET matrix elements $\langle O_{i, \, \mu} \rangle^{(1)}$
can be further written as \cite{Beneke:2005vv}
\begin{eqnarray}
\langle O_{i, \, \mu} \rangle^{(1)} =
\sum_j \, \left [M_{i j, \, \rm bare}^{(1), \, R}  + Z_{i j}^{(1)} \right ] \,
\langle O_{j, \, \mu} \rangle^{(0)} \,,
\label{SCET matrix element at NLO}
\end{eqnarray}
where $M_{i j, \, \rm bare}^{(1), \, R}$ are the bare matrix elements dependent on the IR regularization scheme
and $Z_{i j}^{(1)}$ are the UV renormalization constants at one loop.
When both UV and IR divergences are coped with dimensional regularization, the bare SCET matrix elements vanish
due to the resulting scaleless integrals from the corresponding one-loop diagrams.
Comparing the coefficients of $\langle O_{i, \, \mu} \rangle^{(0)} \, (i=1, 2) $ on
both sides of (\ref{matching relation at NLO}) with the aid of (\ref{SCET matrix element at NLO}) yields
\begin{eqnarray}
C_i^{(1)} &=& T_i^{(1)} - \sum_{j=1, 2, E} \, C_j^{(0)} \ast Z_{j i}^{(1)} \,, 
\label{general NLO matching condition}
\end{eqnarray}
The SCET operators $O_{1, \mu}$ and $O_{2, \mu}$ do not mix into each other,
which can be verified explicitly
by computing the one-loop correction to the two SCET matrix elements
\begin{eqnarray}
\langle O_{i, \, \mu} \rangle^{(1)} = Z_{ii}^{(1)} \, \langle O_{i, \, \mu} \rangle^{(0)} \,, \,\,
{\rm with} \,\,\, i=1, 2 \,.  
\end{eqnarray}
The collinear subtraction term $Z_{ii}^{(1)} \, \langle O_{i, \, \mu} \rangle^{(0)}$ and
the UV renormalization  of the QCD pseudoscalar current $\bar b \, \gamma_5 \, u$
will remove the divergent terms of the NLO QCD amplitude $T_{i}^{(1)}$ to guarantee
that the perturbative matching coefficients entering the factorization
formulae of the correlation function (\ref{def: correlation function}) are free of singularities and
are entirely stemmed from integrating out the hard-scale dynamics of $\Pi_{\mu}$.
We further turn to determine the IR subtraction term $Z_{E i}^{(1)} \, (i=1, 2)$ originated from
the renormalization mixing of the evanescenet operators $O_{E, \, \mu}$ into the physical SCET
operators $O_{1, \, \mu}$ and $O_{2, \, \mu}$. As discussed in \cite{Buras:1989xd,Beneke:2005vv,Dugan:1990df},
the renormalization constants $Z_{E i}^{(1)} \, (i=1, 2)$  will be determined by implementing
the prescription that the IR finite matrix element of the evanescent operator $O_{E, \, \mu}$ vanishes,
when applying dimensional regularization only to the UV divergences and regularizing the IR singularities
with any other scheme different from the dimensions of  spacetime.
In accordance with (\ref{SCET matrix element at NLO}) this amounts to
\begin{eqnarray}
Z_{E i}^{(1)} = - M_{E i, \, \rm bare}^{(1), \, \rm off} \,.
\label{IR subtrsaction ZEi}
\end{eqnarray}
Inserting (\ref{IR subtrsaction ZEi}) into (\ref{general NLO matching condition}) leads to
the following master formula
\begin{eqnarray}
C_{i}^{(1)} = T_i^{(1)} - C_i^{(0)} \ast Z_{ii}^{(1)} +  C_E^{(0)} \ast M_{E i, \, \rm bare}^{(1), \, \rm off}
=  T_{i, \, \rm hard}^{(1), \, \rm reg} +  C_E^{(0)} \ast M_{E i, \, \rm bare}^{(1), \, \rm off}  \,,
\label{master formula for the twist 2}
\end{eqnarray}
where $T_{i, \, \rm hard}^{(1), \, \rm reg}$ is the regularized terms of the NLO
hard contribution to the QCD matrix  element $F_{\mu}$ as presented in (\ref{NLO hard contribution})
and $i=1, 2$.

\begin{figure}
\begin{center}
\includegraphics[width=0.8 \columnwidth]{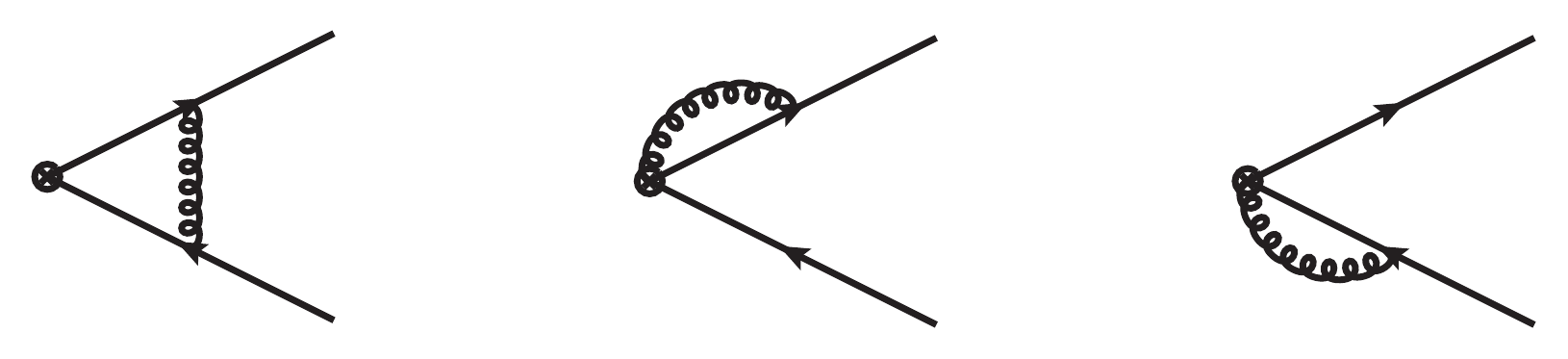} \\
(a) \hspace{4.5 cm} (b) \hspace{4.5 cm} (c) \\
\vspace*{0.1cm}
\caption{The one-loop diagrams for the SCET matrix element $\langle O_{E, \, \mu} \rangle$. }
\label{fig:evanescent operator effect at NLO}
\end{center}
\end{figure}

We proceed to compute the one-loop matrix element of the evanescent SCET operator $\langle O_{E, \, \mu} \rangle^{(1)}$
by evaluating the effective diagrams shown in figure \ref{fig:evanescent operator effect at NLO}.
Employing the SCET Feynman rules, we find that only the diagram (a) with a collinear-gluon exchange between two
collinear quarks could give rise to a non-trivial contribution to $M_{E i, \, \rm bare}^{(1), \, \rm off}$.
Evaluating this one-loop SCET diagram explicitly yields
\begin{eqnarray}
\langle O_{E, \, \mu} (z, z^{\prime})  \rangle^{(1)} & \supset &
- i \, g_s^2 \, C_F \, \int {d^D l \over (2 \, \pi)^D} \,
{1 \over [(z \, p+l)^2 + i 0] [(l- \bar z \, p)^2+i0] [l^2+ i 0]} \nonumber \\
&& \bar u(z \, p) \, \gamma_{\nu \perp} \,  \not l_{\perp} \, \Gamma_3 \,
\not l_{\perp} \, \gamma_{\perp}^{\nu} \, v(z \, p) \,
\delta \left (z^{\prime} - z - {n \cdot l \over n \cdot p} \right )  \,,
\end{eqnarray}
which  only  generates a non-vanishing contribution proportional to the SCET
matrix element $\langle O_{E, \, \mu} \rangle^{(0)}$ at ${\cal O}(\epsilon)$
with the NDR scheme of $\gamma_5$.
Explicitly, we obtain
\begin{eqnarray}
 C_E^{(0)} \ast M_{E i, \, \rm bare}^{(1), \, \rm off} = 0 \,, \,\,
{\rm with} \,\,\, i=1, 2 \,,
\end{eqnarray}
from which the one-loop hard matching coefficients can be written as
\begin{eqnarray}
C_{i}^{(1)} = T_{i, \, \rm hard}^{(1), \, \rm reg} \,.
\end{eqnarray}

Now we are ready to demonstrate the factorization-scale independence of the factorization formula
for the vacuum-to-photon correlation function (\ref{def: correlation function})
\begin{eqnarray}
\Pi_{\mu}(p, q) &=& {i  \over 2} \, g_{\rm em} \, Q_u \,
\chi(\mu) \, \langle \bar q q \rangle(\mu) \,\epsilon^{\ast \, \alpha}(p)
\left [ g_{\mu \alpha}^{\perp} - i \, \epsilon_{\mu \alpha \nu \beta} \,
n^{\nu} \, v^{\beta} \right ] \,  \int_0^1 \, d z \, \phi_{\gamma}(z \,, \mu) \,\, \nonumber \\
&& \frac{n \cdot p \,\, \bar n \cdot q}{z (p+q)^2 + \bar z  \, q^2 - m_b^2 + i 0} \,
\left [ 1 +{C_{i}^{(1)}(z, (p+q)^2, q^2)  \over C_{i}^{(0)}(z, (p+q)^2, q^2) }  \right ] +{\cal O}(\alpha_s^2)  \,.
\label{NLO factorization formula at LT}
\end{eqnarray}
To this end, we need to make use of the evolution equation for the leading-twist photon DA
\begin{eqnarray}
\mu^2 \, {d \over d \mu^2} \, \left[ \chi(\mu) \, \langle \bar q q \rangle (\mu) \,
 \phi_{\gamma} (z, \mu) \right ]
= \int_0^1 \, d z^{\prime} \, \widetilde{V}(z, z^{\prime}) \,
\left[ \chi(\mu) \, \langle \bar q q \rangle (\mu) \,
\phi_{\gamma} (z^{\prime}, \mu) \right ] \,,
\label{RGE of the twist-2 photon  DA}
\end{eqnarray}
with the renormalization kernel $\widetilde{V}(z, z^{\prime})$ expanded perturbatively in QCD
\begin{eqnarray}
\widetilde{V}(z, z^{\prime}) = \sum_{n=0} \, \left ( {\alpha_s \over 4 \, \pi} \right )^{n+1} \,
 \widetilde{V}_n(z, z^{\prime})\,,
\end{eqnarray}
and the RG equation for the bottom-quark mass \cite{Chetyrkin:1997dh,Vermaseren:1997fq}
\begin{eqnarray}
{d \, m_b(\mu)\over d \ln \mu} = - \sum_{n=0} \, \left ( {\alpha_s(\mu) \over 4 \pi} \right )^{n+1} \, \gamma_m^{(n)}  \,,
\qquad \gamma_m^{(0)}  = 6 \, C_F \,.
\end{eqnarray}
The explicit expression of the one-loop evolution kernel $\widetilde{V}_0$ is given by \cite{Lepage:1979zb,Shifman:1980dk}
\begin{eqnarray}
\widetilde{V}_0(z, z^{\prime}) =  2 \, C_F \, \left [ {\bar z \over \bar z^{\prime}}
 {1 \over z - z^{\prime}} \, \theta(z - z^{\prime})
 + {z \over z^{\prime}} \, {1 \over z^{\prime} - z} \,
 \, \theta(z^{\prime} - z)  \right ]_{+} - C_F \, \delta(z - z^{\prime})\,,
\end{eqnarray}
where the plus function is defined as
\begin{eqnarray}
\left [ f(z, z^{\prime}) \right ]_{+} = f(z, z^{\prime})
- \delta(z-z^{\prime}) \, \int_0^1 d t \, f(t, z^{\prime}) \,.
\end{eqnarray}
It is then straightforward to write down
\begin{eqnarray}
{d \Pi_{\mu}(p, q) \over d \ln \mu} &=&  {i  \over 2} \, g_{\rm em} \, Q_u \,
\chi(\mu) \, \langle \bar q q \rangle(\mu) \,\epsilon^{\ast \, \alpha}(p)
\left [ g_{\mu \alpha}^{\perp} - i \, \epsilon_{\mu \alpha \nu \beta} \,
n^{\nu} \, v^{\beta} \right ] \,  \int_0^1 \, d z \, \phi_{\gamma}(z \,, \mu) \,\, \nonumber \\
&& \frac{n \cdot p \,\, \bar n \cdot q}{z (p+q)^2 + \bar z  \, q^2 - m_b^2 + i 0} \,
\left \{ {\alpha_s \, C_F \over 4 \, \pi} \, 6 + {\cal O}(\alpha_s^2)\right \}  \,.
\end{eqnarray}
The residual $\mu$ dependence at one loop arises from the UV renormalization of the pseudoscalar
QCD current defining the correlation function (\ref{def: correlation function}).
Distinguishing the renormalization scale $\mu$, due to the non-conservation of the
pseudoscalar current in QCD, from the factorization scale $\mu$ governing the RG evolution
in SCET, we are led to conclude that the factorization formula (\ref{NLO factorization formula at LT})
of $\Pi_{\mu}(p, q)$ is indeed independent of the scale $\mu$ at one-loop accuracy.

 According to the QCD factorization formula (\ref{NLO factorization formula at LT})
 for the correlation function $\Pi_{\mu}(p, q)$, we cannot avoid the parametrically large logarithms
 of  ${\cal O} (\ln{(m_b^2 / \Lambda^2)})$ by adopting a universal scale $\mu$ in the hard matching coefficient
 and in the photon DA. We will perform resummmation of the above-mentioned large logarithms at NLL accuracy
 by applying the two-loop RG equation of the twist-two photon DA and by setting the factorization scale as
 $\mu \sim m_b$. The NLO evolution kernel $\widetilde{V}_1$ 
 in QCD can be decomposed as follows \cite{Belitsky:1999gu,Mikhailov:2008my,Belitsky:2000yn}
 \begin{eqnarray}
\widetilde{V}_1(z, z^{\prime}) = {N_f  \over 2} \, C_F \, \widetilde{V}_N(z, z^{\prime})
+  C_F \, C_A \, \widetilde{V}_G(z, z^{\prime})
+ C_F^2 \, \widetilde{V}_F(z, z^{\prime})\,,
\end{eqnarray}
 where the explicit expressions of the evolution functions can be found in \cite{Mikhailov:2008my}.
 Symmetry properties of the RG evolution equation (\ref{RGE of the twist-2 photon  DA}) imply the
 series expansion of the leading-twist photon DA in terms of the Gegenbauer polynomials
 \begin{eqnarray}
 \phi_{\gamma}(z, \mu)= 6 \, z \, \bar z \, \sum_{n=0}^{\infty} \, a_n(\mu) \, C_n^{3/2}(2 z -1)\,.
 \label{Gegenbauer expansion}
\end{eqnarray}
The two-loop evolution of the Gegenbauer moment $a_n(\mu)$ can then be obtained as follows
\begin{eqnarray}
\chi(\mu) \, \langle \bar q q \rangle (\mu) \, a_n(\mu) &=&
E_{T, n}^{\rm NLO}(\mu, \mu_0) \, \chi(\mu_0) \, \langle \bar q q \rangle (\mu_0) \, a_n(\mu_0) \nonumber \\
&& +  {\alpha_s(\mu) \over 4 \pi} \, \sum_{k=0}^{n-2} \,  E_{T, n}^{\rm LO}(\mu, \mu_0) \,
d_{T, n}^{k}(\mu, \mu_0) \, \chi(\mu_0) \, \langle \bar q q \rangle (\mu_0) \, a_n(\mu_0)   \,,
\label{two loop evolution of the photon  moments}
\end{eqnarray}
where $k, n=0, 2, 4, ...$ and the explicit expressions of the RG functions $E_{T, n}^{\rm (N)LO}$
and the off-diagonal mixing coefficients can be found in Appendix A of \cite{Wang:2017ijn}.
In contrast to the LO evolution in QCD, the Gegenbauer coefficients $a_n(\mu)$ do not renormalize
multiplicatively at NLO accuracy. Inserting  (\ref{Gegenbauer expansion}) and
(\ref{two loop evolution of the photon  moments}) into the NLO factorization formula
(\ref{NLO factorization formula at LT}) gives rise to the NLL resummation improved expression
\begin{eqnarray}
\Pi_{\mu}(p, q) &=& g_{\rm em} \, Q_u \,n \cdot p \,
\chi(\mu) \, \langle \bar q q \rangle(\mu) \,\epsilon^{\ast \, \alpha}(p)
\left [ g_{\mu \alpha}^{\perp} - i \, \epsilon_{\mu \alpha \nu \beta} \,
n^{\nu} \, v^{\beta} \right ]  \nonumber \\
&& \, \times \sum_{n=0} \, a_n(\mu) \,\,  K_n((p+q)^2, q^2)\,
+ {\cal O}(\alpha_s^2)  \,,
\label{NLL factorization formula at LT}
\end{eqnarray}
where the perturbative function $K_n((p+q)^2, q^2)$ is determined by
\begin{eqnarray}
K_n=\int_0^1 \, d z \, \left [C_{i}^{(0)}(z, (p+q)^2, q^2)  + C_{i}^{(1)}(z, (p+q)^2, q^2)  \right ] \,
\left [ 6 \, z \, \bar z \, C_{n}^{3/2}(2 \, z -1) \right ] \,.
\end{eqnarray}

To construct the sum rules for the twist-two hadronic photon correction to the $B \to \gamma \ell \nu$ form factors,
we need to derive the dispersion representation for the NLL factorization formula (\ref{NLL factorization formula at LT}).
Applying the spectral representations of the convolution integrals collected in Appendix
\ref{app:spectral resp}, we can readily obtain
\begin{eqnarray}
\Pi_{\mu}(p, q) &=& {i \over 2} \, g_{\rm em} \, Q_u \, n \cdot p \,\, \bar n \cdot q \,
\chi(\mu) \, \langle \bar q q \rangle(\mu) \,\epsilon^{\ast \, \alpha}(p)
\left [ g_{\mu \alpha}^{\perp} - i \, \epsilon_{\mu \alpha \nu \beta} \,
n^{\nu} \, v^{\beta} \right ]  \nonumber \\
&& \times \int_0^{\infty} \, {d s \over s-(p+q)^2- i 0} \,
\left [\rho^{(0)}(s, q^2) +{\alpha_s \, C_F \over 4 \, \pi} \, \rho^{(1)}(s, q^2) \right ] \,,
\label{NLL factorization formula for the twist-two correction}
\end{eqnarray}
where the LO spectral function $\rho^{(0)}(s, q^2)$ is given by
\begin{eqnarray}
\rho^{(0)}(s, q^2)= - {1 \over s-q^2} \,\, \phi_{\gamma} \left ({m_b^2 - q^2 \over s-q^2}, \mu \right ) \,
\theta(s-m_b^2)\,.
\end{eqnarray}
The resulting NLO spectral function $\rho^{(1)}(s, q^2)$ is rather involved and can be written as
\begin{eqnarray}
\rho^{(1)}(s, q^2) &=& (-2) \, \ln \left ( {\mu^2 \over m_b^2} \right ) \, {1 \over r_3-r_2} \,
\bigg \{ \int_0^1 \, d z \, \bigg [ \ln \left ( {z \, r_3 + \bar z \, r_2 - 1 \over 1 - r_2} \right ) \,
\theta(z \, r_3 + \bar z \, r_2 - 1) \nonumber \\
&& + \, \ln \bigg |{ z \, r_3 + \bar z \, r_2 - 1 \over r_3-1} \bigg | \,
\left ( \theta(z \, r_3 + \bar z \, r_2 -1) - \theta(r_3-1)\right ) \bigg  ] \, \phi_{\gamma}^{\prime}(z, \mu) \,
 \nonumber \\
&& + \, \int_0^1 \, d z \, \left [ {\theta(z \, r_3 + \bar z \, r_2 -1) \over z}
- {\theta(z \, r_3 + \bar z \, r_2 -1) - \theta(r_3-1) \over \bar z}\right ] \,  \phi_{\gamma}(z, \mu)  \nonumber \\
&& + \, {3 \over r_3 -r_2} \, \theta(r_3-1) \, \phi_{\gamma}^{\prime} \left ({1- r_2 \over r_3-r_2}, \mu \right )
 - {1 \over r_3 -r_2} \, {2 \, \pi^2 \over 3} \, \phi_{\gamma} \left ( {1- r_2 \over r_3-r_2}, \mu \right ) \,
 \theta(r_3-1)\, \bigg \} \nonumber \\
&&  + \, 2 \, \int_0^1 \, d z \, \phi_{\gamma}(z, \mu) \,
\bigg \{ \left [ {1 \over z \, r_3 + \bar z \, r_2 -1} - {1 \over z} \, {1 \over r_3 -r_2} \right ]  \,
\ln \left ( 1 +  { z \, r_3 + \bar z \, r_2 -1 \over 1-r_2}  \right )  \nonumber \\
&& \hspace{0.3 cm} \times \, \theta(z \, r_3 + \bar z \, r_2 -1)
- \left [ {1 \over z \, r_3 + \bar z \, r_2 -1} + {1 \over \bar z} \, {1 \over r_3 -r_2}  \right ] \,
\ln \left ( 1 +  { z \, r_3 + \bar z \, r_2 -1 \over 1-r_3}  \right )  \nonumber \\
&& \hspace{0.3 cm} \times \, \theta \left( { z \, r_3 + \bar z \, r_2 -1 \over 1-r_3} \right ) \bigg \}
+ \int_0^1 \, d z \, {2 \over r_3 -r_2} \, \left [ \ln^2 ( z \, r_3 + \bar z \, r_2 -1) - {\pi^2 \over 3} \right ] \, \nonumber \\
&&  \hspace{0.3 cm} \times \, \theta(z \, r_3 + \bar z \, r_2 -1) \, \, \phi_{\gamma}^{\prime}(z, \mu)
+ \int_0^1 \, d z \, \phi_{\gamma}(z, \mu)  \,  \left ({1 \over z}  - {1 \over \bar z} \right ) \, \nonumber \\
&&   \hspace{0.3 cm} \times \, \bigg [ {2 \over r_3 -r_2} \, \ln (z \, r_3 + \bar z \, r_2 -1) \, \theta(z \, r_3 + \bar z \, r_2 -1)
+ \, \delta(r_3-r_2) \, \ln^2(1-r_2) \bigg ]  \nonumber \\
&& + \, {1 \over r_3-r_2} \, \phi_{\gamma} \left ( {1-r_2 \over r_3-r_2}, \mu  \right ) \,
\theta(r_3-1)  \, \left [ \ln^2 (1-r_2) + \ln^2 (r_3-1) - \pi^2 \right ] \nonumber \\
&& + \, 2 \, \int_0^1 d z \,\phi_{\gamma}(z, \mu) \, \left [ {\cal P}{1 \over z \, r_3 + \bar z \, r_2 -1}
- {1 \over \bar z} \, {1 \over r_2 -r_3} \right ] \, \ln (r_3 -1) \, \theta(r_3 -1) \nonumber \\
&& + \, {\theta(r_3-1) \over r_3-r_2} \, \left [{4 \over r_3} \, \ln (r_3-1)
+ {(2-r_2) \, (8 - 3 \, r_2)  \over r_2 \, (4 - r_2)} \, \ln (1-r_2)  \right ] \,
\phi_{\gamma} \left( {1-r_2 \over r_3-r_2}, \mu \right )  \nonumber \\
&& + \, 3 \,  {\theta(r_3-1) \over (r_3-r_2)^2}  \,
\left [ \phi_{\gamma}^{\prime} \left( {1-r_2 \over r_3-r_2}, \mu \right )
+ 2 \, \ln (r_3-1) \, \phi_{\gamma}^{\prime} \left(z=1, \mu \right ) \right ] \nonumber \\
&& - \, \int_0^1 \, d z \, \ln (z \, r_3 + \bar z \, r_2 -1) \,
{ \theta(z \, r_3 + \bar z \, r_2 -1) \over r_3 -r_2}\, \left [ {6 \over r_3-r_2} \, {d^2 \over dz^2}
+ {r_2 \over 1-r_2} \, {d \over dz} \right ] \, \phi_{\gamma}(z, \mu) \nonumber \\
&& - \, \theta(r_3-1) \, \int_0^1 d z \, \phi_{\gamma}(z, \mu) \, \bigg \{ \theta \left (z - {1- r_2 \over r_3-r_2} \right ) \,
\bigg [ { r_2 (r_2-2) (1+z) + 2 \, z \over r_2 \, (1-r_2) \, z \bar z} \, {1 \over r_2 - r_3}  \nonumber \\
&& +  \,  {2 \, (z - 2 \, \bar z \, r_2)  \over \bar z \, r_2} \, {1 \over z \, r_3 + \bar z \, r_2}  \bigg ]
+ {2 \, (1-r_2) \over \bar z \, r_2} \, {1 \over r_3 -r_2}
- {2 \, (1 - 3\, \bar z \, r_2) \over \bar z\, r_2 \, (1- \bar z \, r_2)} \, {1 \over r_3} \nonumber \\
&& - { 4 \, z \over 1- \bar z \, r_2} \, {\cal P}{1 \over z \, r_3 + \bar z \, r_2 -1}  \bigg \}
+ {r_2 \over 1-r_2} \, \delta(r_2 -r_3) \, \int_0^1 \, dz \, {\phi_{\gamma}(z, \mu) \over z}  \nonumber \\
&& - {1 \over r_3 -r_2} \, {16 \, r_2 -15 \over 2 \, (1-r_2)} \,
\phi_{\gamma}  \left ( {1 - r_2 \over r_3 -r_2 }  , \mu \right ) \, \theta(r_3-1)\,,
\end{eqnarray}
where ${\cal P}$ indicates the principle-value prescription. Finally, the NLL sum rules for
the hadronic  photon correction to the $B \to \gamma$ form factors at leading twist
can be written as
\begin{eqnarray}
{f_B\, m_B \over m_b+m_u} \, F_{V, \, \rm photon}^{\rm 2PLT}(n \cdot p)
&=&{f_B\, m_B \over m_b+m_u} \, F_{A, \, \rm photon}^{\rm 2PLT}(n \cdot p) \nonumber \\
&=& - \, Q_u \, \chi(\mu) \, \langle \bar q q \rangle(\mu) \,\, \bar n \cdot q  \,
\int_{0}^{s_0} \, ds \, {\rm exp}
\left [ - {s -m_B^2 \over M^2} \right ] \, \nonumber \\
&& \, \times \left [\rho^{(0)}(s, q^2) +{\alpha_s \, C_F \over 4 \, \pi} \, \rho^{(1)}(s, q^2) \right ]
+ \, {\cal O}(\alpha_s^2)\,.
\label{NLL sum rule at twist-2}
\end{eqnarray}
It is evident that the twist-two hadronic photon correction preserves the symmetry relation of
the two form factors $F_V$ and $F_A$ at leading power in $\Lambda/m_b$.

\section{Higher-twist hadronic photon corrections in QCD}
\label{sect:higher-twist-photon}

In this section we will aim at computing the higher-twist hadronic photon corrections
to the $B \to \gamma \ell \nu$ decay form factors at LO in $\alpha_s$, up to the twist-four accuracy,
from the LCSR approach. Following the discussion on a general classification of the photon DAs \cite{Ball:2002ps},
we will need to take into account the subleading-power contributions arising from the
light-cone matrix elements of both the two-body and three-body  collinear operators.
To achieve this goal, we first demonstrate  QCD factorization for
the two-particle and three-particle higher-twist contributions to the vacuum-to-photon
correlation function (\ref{def: correlation function}) and then construct the tree-level sum rules
for the form factors $F_V$ and $F_A$ following the standard strategy.

\subsection{Higher-twist two-particle corrections}

Employing the light-cone expansion of the bottom-quark propagator
and keeping the subleading-power contributions to
the correlation function  (\ref{def: correlation function}) leads to
\begin{eqnarray}
\Pi_{\mu}(p, q) & \supset & \int {d^4 k \over (2 \pi)^4} \, \int d^4 x \,
e^{i \, (q-k) \cdot x} \, {k^{\nu} \over k^2 - m_b^2} \,
\langle \gamma(p)| \bar u(x) \, \sigma_{\mu \nu} \, (1 + \gamma_5)  \, u(0)| 0 \rangle \nonumber \\
&&  - \, i \, \int {d^4 k \over (2 \pi)^4} \, \int d^4 x \,
e^{i \, (q-k) \cdot x} \, {m_b \over k^2 - m_b^2} \,
\langle \gamma(p)| \bar u(x) \, \gamma_{\mu} \, (1 - \gamma_5)  \, u(0)| 0 \rangle \,.
\end{eqnarray}
Making use of the definitions of the higher-twist photon DAs displayed in Appendix \ref{app:Higher-twist photon DAs},
it is straightforward to  write down
\begin{eqnarray}
\Pi_{\mu}(p, q) & \supset & {i \over 4}  \, g_{\rm em} \, Q_q \, (p \cdot q) \,
\int_0^1 d z \, \bigg \{\epsilon_{\mu}^{\ast} \,
\left [ {\rho_{A, 2}^{\rm 2PHT}((p+q)^2, q^2, z) \over [(z \, p + q)^2 - m_b^2 + i \, 0]^2}
+ {\rho_{A, 3}^{\rm 2PHT}((p+q)^2, q^2, z) \over [(z \, p + q)^2 - m_b^2 + i \, 0]^3}   \right ]  \nonumber \\
&& - \, i \, \varepsilon_{\mu \nu \alpha \beta} \, \epsilon^{\ast \, \nu} \, n^{\alpha} \, v^{\beta} \,
\left [ {\rho_{V, 2}^{\rm 2PHT}((p+q)^2, q^2, z) \over [(z \, p + q)^2 - m_b^2 + i \, 0]^2}
+ {\rho_{V, 3}^{\rm 2PHT}((p+q)^2, q^2, z) \over [(z \, p + q)^2 - m_b^2 + i \, 0]^3}   \right ]  \bigg \}  \,,
\end{eqnarray}
where the explicit expressions of the invariant functions $\rho_{V(A), \, i}^{\rm 2PHT} \,\, (i=2 \,, 3)$
are given by
\begin{eqnarray}
\rho_{V, 2}^{\rm 2PHT}((p+q)^2, q^2, z) &=& 2 \, m_b \, f_{3 \gamma}(\mu) \, \psi^{(a)}(z, \mu)
- \langle \bar q q \rangle(\mu)  \,\, \mathbb{A}(z, \mu)\,,  \nonumber \\
\rho_{V, 3}^{\rm 2PHT}((p+q)^2, q^2, z) &=& - 2 \, m_b^2 \,\,
\langle \bar q q \rangle(\mu)  \,\, \mathbb{A}(z, \mu) \,,  \nonumber \\
\rho_{A, 2}^{\rm 2PHT}((p+q)^2, q^2, z) &=& 4 \, m_b \, f_{3 \gamma}(\mu) \,
\bar{\psi}^{(v)}(z, \mu) +  \left [  \mathbb{A}(z, \mu) - 2 \, \bar{h}_{\gamma}(z, \mu) \right ] \,
\langle \bar q q \rangle(\mu) \,,  \nonumber \\
\rho_{A, 3}^{\rm 2PHT}((p+q)^2, q^2, z) &=& - 2 \, m_b^2 \, \langle \bar q q \rangle(\mu) \,
\left [ \mathbb{A}(z, \mu) - \, 2 \, \bar{h}_{\gamma}(z, \mu)  \right ]\,.
\label{correlator:2PHT effect}
\end{eqnarray}
The two new functions $\bar{\psi}^{(v)}(z, \mu)$
and $\bar{h}_{\gamma}(z, \mu)$ introduced in (\ref{correlator:2PHT effect}) are defined by
\begin{eqnarray}
\bar{\psi}^{(v)}(z, \mu) =  2 \, \int_0^z \, d \alpha \, \psi^{(v)}(\alpha, \mu)\,,  \qquad
\bar{h}_{\gamma}(z, \mu) = - \, 4\, \int_0^z \, d \alpha (z - \alpha) \, h_{\gamma}(\alpha, \mu) \,.
\end{eqnarray}
The resulting LCSR for the two-particle higher-twist hadronic photon corrections to the $B \to \gamma \ell \nu$
form factors can be further derived as follows
\begin{eqnarray}
&& - {f_B\, m_B \over m_b+m_u} \, {\rm exp} \left ( - {m_B^2 \over M^2 } \right ) \,
 F_{V(A), \, \rm photon}^{\rm 2PHT, \, LL}(n \cdot p) \nonumber \\
&& = {Q_q \over 4} \, \bigg \{ {1 \over m_b^2-q^2} \, {\rm exp} \left ( - {s_0 \over M^2 } \right )  \,
\rho_{V(A), 2}^{\rm 2PHT} \left (s_0, q^2, z=f_1(s_0, q^2) \right ) \nonumber \\
&& \hspace{0.4 cm} + \, \int_{m_b^2}^{s_0} \, {d s \over m_b^2-q^2} \, {1 \over M^2} \,
\, {\rm exp} \left ( - {s \over M^2 } \right ) \,
\rho_{V(A), 2}^{\rm 2PHT} \left (s, q^2, z=f_1(s, q^2) \right ) \nonumber \\
&& \hspace{0.4 cm} + \, {1 \over (s_0-q^2)^2}  \, {\rm exp} \left ( - {s_0 \over M^2 } \right )  \,
{d \over d z} \, \left [{1 \over 2 \, z} \, \rho_{V(A), 3}^{\rm 2PHT}(s_0, q^2, z)  \right ] \bigg|_{z=f_1(s_0, q^2)} \nonumber \\
&&  \hspace{0.4 cm} + \,  {s_0-q^2 \over 2 \, (m_b^2 -q^2)^2} \,
{d \over d s_0} \, \left [{\rm exp} \left ( - {s_0 \over M^2 }  \right ) \,
\rho_{V(A), 3}^{\rm 2PHT}(s_0, q^2, z) \right ] \bigg|_{z=f_1(s_0, q^2)} \, \nonumber \\
&&   \hspace{0.4 cm} - \, \, \int_{m_b^2}^{s_0} \,  ds \, {s-q^2 \over 2 \, (m_b^2-q^2)^2} \,
{d^2 \over d s^2} \, \left [{\rm exp} \left ( - {s \over M^2 }  \right ) \,
\rho_{V(A), 3}^{\rm 2PHT}(s, q^2, z) \right ] \bigg|_{z=f_1(s, q^2)}  \bigg \} \,,
\label{LO sum rule for the 2PHT}
\end{eqnarray}
where we have defined $f_1(s, q^2)= {(m_b^2 -q^2) / (s-q^2)}$  to compactify the above expressions.

Several comments on the tree-level sum rules for the higher-twist corrections to the form factors $F_V$
and $F_A$ presented in (\ref{LO sum rule for the 2PHT}) are in order.
\begin{itemize}

\item{It is evident from (\ref{correlator:2PHT effect}) that the higher-twist two-particle hadronic photon
corrections can lead to the symmetry-breaking contributions to  the $B \to \gamma \ell \nu$ form factors
already at tree level, in agreement with the observation made in \cite{Ball:2003fq}.
However, it needs to be pointed out that the subleading-twist effects do not always violate
the symmetry relation of the two $B \to \gamma$ form factors at leading power in $\Lambda/m_b$
\cite{Wang:2016qii}.}

\item{Applying the power-counting scheme for the threshold parameter (\ref{power counting for the threshold})
and the end-point behaviours of the two-particle photon DAs $\bar{\psi}^{(v)}(z, \mu)$, $\psi^{(a)}(z, \mu)$,
$\mathbb{A}(z, \mu)$  and $\bar{h}_{\gamma}(z, \mu)$, we can readily identify the heavy-quark scaling
for the two-particle higher-twist corrections
\begin{eqnarray}
F_{V, \, \rm photon}^{\rm 2PHT, \, LL}(n \cdot p) \sim
F_{A, \, \rm photon}^{\rm 2PHT, \, LL}(n \cdot p) \sim \left ({\Lambda \over m_b} \right )^{3/2} \,,
\end{eqnarray}
which is of the same power as the twist-two hadronic photon contribution obtained
in (\ref{power counting of the twist-2 hadronic photon contribution}) and is
suppressed by only one factor of $\Lambda/m_b$ compared with the direct photon contribution
(\ref{power counting of the LP contribution}).  We are then led to conclude that there is generally no
correspondence between the heavy-quark expansion and the twist expansion for the $B \to \gamma \ell \nu$
form factors in the LCSR approach (see \cite{Agaev:2010aq} for a discussion in the context of
the pion-photon form factor).}

\end{itemize}

\subsection{Higher-twist three-particle corrections}

\begin{figure}
\begin{center}
\includegraphics[width=0.4 \columnwidth]{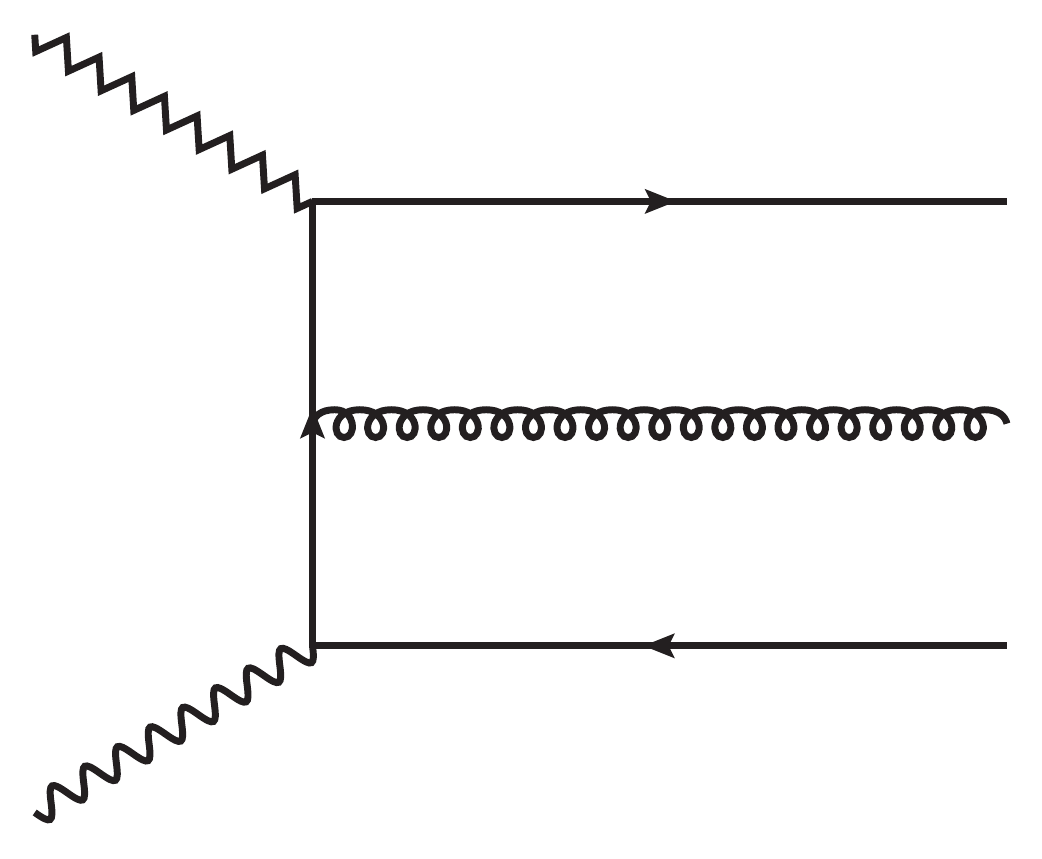} \\
\vspace*{0.1cm}
\caption{Diagrammatical representation of the three-particle contribution to
the QCD amplitude $F_{\mu}(p, q)$ defined in (\ref{QCD amplitude at twist-2})
at tree level. }
\label{fig:higher-twist 3-particle correlator at LO}
\end{center}
\end{figure}

We will proceed to compute the higher-twist three-particle hadronic photon corrections
to the $B \to \gamma \ell \nu$ form factors at tree level with the sum rule technique.
Following the standard strategy, we first compute the three-particle contribution to
the four-point QCD amplitude $F_{\mu}(p, q)$  (\ref{QCD amplitude at twist-2})  displayed
in figure \ref{fig:higher-twist 3-particle correlator at LO}.
Keeping the one-gluon part for the light-cone expansion of the bottom-quark propagator
in the background gluon/photon field \cite{Balitsky:1987bk,Belyaev:1994zk}
\begin{eqnarray}
&& \langle 0 | T \{\bar b(x), \, b(0)  \}| 0 \rangle  \nonumber  \\
&& \supset i \, g_s \, \int {d^4 k \over (2 \, \pi)^4} \, e^{- i \, k \cdot x} \,
\int_0^1 d v \, \left [ {v \, x_{\mu} \over k^2-m_b^2 } \, G^{\mu \nu}(v x) \, \gamma_{\nu}
- {\not \! k + m_b \over 2(k^2-m_b^2)^2} \, G^{\mu \nu}(v x) \, \sigma_{\mu \nu}\right ]  \nonumber \\
&& \hspace{0.3 cm} + \,  i \, g_{\rm em} \, Q_b \, \int {d^4 k \over (2 \, \pi)^4} \, e^{- i \, k \cdot x} \,
\int_0^1 d v \, \left [ {v \, x_{\mu} \over k^2-m_b^2 } \, F^{\mu \nu}(v x) \, \gamma_{\nu}
- {\not \! k + m_b \over 2(k^2-m_b^2)^2} \, F^{\mu \nu}(v x) \, \sigma_{\mu \nu}\right ], \hspace{0.3 cm}
\end{eqnarray}
and employing the definitions of the three-particle photon DAs in Appendix \ref{app:Higher-twist photon DAs},
we obtain
\begin{eqnarray}
\Pi_{\mu}(p, q) & \supset & i \, g_{\rm em} \, Q_q \, (p \cdot q) \,
\int_0^1 d v \, \int [{\cal D} \alpha_i] \, \bigg \{\epsilon_{\mu}^{\ast} \,
\bigg  [ {\rho_{A, 2}^{\rm 3P}((p+q)^2, q^2, \alpha_i, v) \over
[((\alpha_q + v \, \alpha_g) \, p + q)^2 - m_b^2 + i \, 0]^2} \nonumber \\
&& + {\rho_{A, 3}^{\rm 3P}((p+q)^2, q^2, \alpha_i, v) \over
[((\alpha_q + v \, \alpha_g) \, p + q)^2 - m_b^2 + i \, 0]^3}   \bigg ]
- \, i \, \varepsilon_{\mu \nu \alpha \beta} \,
\epsilon^{\ast \, \nu} \, n^{\alpha} \, v^{\beta} \, \nonumber \\
&&  \times \left [ {\rho_{V, 2}^{\rm 3P}((p+q)^2, q^2, \alpha_i, v)
\over [((\alpha_q + v \, \alpha_g) \, p + q)^2 - m_b^2 + i \, 0]^2}
+ {\rho_{V, 3}^{\rm 3P}((p+q)^2, q^2, \alpha_i, v)
\over [((\alpha_q + v \, \alpha_g) \, p + q)^2 - m_b^2 + i \, 0]^3}   \right ]  \bigg \}  \,,
\hspace{0.5 cm}
\end{eqnarray}
where the integration measure is defined as
\begin{eqnarray}
\int [{\cal D} \alpha_i] \equiv \int_0^1 d \alpha_q \, \int_0^1 d \alpha_{\bar q} \,
\int_0^1 d \alpha_g \, \delta \left (1-\alpha_q - \alpha_{\bar q} -\alpha_g \right )\,.
\end{eqnarray}
The resulting expressions for the invariant functions $\rho_{V(A), i}^{\rm 3P} \, (i=2, \, 3)$
are given by
\begin{eqnarray}
\rho_{V, 2}^{\rm 3P}&=&
- \left [S(\alpha_i, \mu) + S_{\gamma}(\alpha_i, \mu) \right ]
 + (1- 2 \, v)\, \widetilde{S}(\alpha_i, \mu)
 - 2\, v \, \left [ T_1(\alpha_i, \mu) -  T_2(\alpha_i, \mu) \right ]  \,, \nonumber \\
\rho_{V, 3}^{\rm 3P} &=& 0  \,, \nonumber \\
\rho_{A, 2}^{\rm 3P}&=& - (1- 2\, v) \, \left [S(\alpha_i, \mu) + S_{\gamma}(\alpha_i, \mu) \right ]
+  \widetilde{S}(\alpha_i, \mu) + T_1(\alpha_i, \mu) + (1- 2\, v) \, T_2(\alpha_i, \mu) \,, \nonumber \\
\rho_{A, 3}^{\rm 3P} &=& 2 \, \left [ (p+q)^2 - q^2 \right ] \,
\left \{ (2 \, v -1) \, \bar{T}_3(\alpha_i, \mu)  -
\left [ \bar{T}_4(\alpha_i, \mu) + \bar{T}_4^{\gamma}(\alpha_i, \mu) \right ] \right \}  \,,
\end{eqnarray}
where we have introduced the following notations
\begin{eqnarray}
\bar{T}_{3(4)}(\alpha_i, \mu) = \int_0^{\alpha_q} \, d \alpha_q^{\prime} \,
T_{3(4)}(\alpha_q^{\prime}, \alpha_{\bar q}, \alpha_g, \mu) \,, \qquad
\bar{T}_4^{\gamma}(\alpha_i, \mu) = \int_0^{\alpha_q} \, d \alpha_q^{\prime} \,
T_4^{\gamma}(\alpha_q^{\prime}, \alpha_{\bar q}, \alpha_g, \mu) \,.
\end{eqnarray}

Implementing the continuum subtraction with the aid of the parton-hadron duality relation
and preforming the Borel transformation in the variable $(p+q)^2 \to s$ gives rise to the desired
sum rules for the three-particle hadronic photon corrections at tree level
\begin{eqnarray}
&& - {f_B\, m_B \over m_b+m_u} \, {\rm exp} \left ( - {m_B^2 \over M^2 } \right ) \,
 F_{V(A), \, \rm photon}^{\rm 3P, \, LL}(n \cdot p) \nonumber \\
&& =  Q_q \, \langle \bar q q \rangle (\mu) \, \bigg \{ \int_{0}^{f_1(s_0, q^2)} d \alpha_q
\,\, \int_{f_2(\alpha_q, s_0, q^2)}^{1-\alpha_q} \, {d \alpha_g \over \alpha_g } \,
{\theta(1-\alpha_q-f_2(\alpha_q, s_0, q^2)) \over m_b^2 -q^2} \,
{\rm exp} \left ( - {s_0 \over M^2 } \right ) \nonumber \\
&&  \hspace{0.4 cm} \times \, \rho_{V(A), 2}^{\rm 3P}
\left (s_0, q^2, \alpha_q, \alpha_{\bar q}, \alpha_g, v={f_2(\alpha_q, s_0, q^2) \over \alpha_g} \right )
+ \int_{m_b^2}^{s_0} \, d s \, \int_{0}^{f_1(s, q^2)} d \alpha_q
\,\, \int_{f_2(\alpha_q, s, q^2)}^{1-\alpha_q} \, {d \alpha_g \over \alpha_g } \, \nonumber \\
&& \hspace{0.4 cm} \times \, {\theta(1-\alpha_q-f_2(\alpha_q, s, q^2)) \over m_b^2 -q^2} \,
{1 \over M^2} \, {\rm exp} \left ( - {s \over M^2 } \right ) \,
\rho_{V(A), 2}^{\rm 3P}
\left (s, q^2, \alpha_q, \alpha_{\bar q}, \alpha_g, v={f_2(\alpha_q, s, q^2) \over \alpha_g} \right )  \nonumber \\
&& \hspace{0.4 cm} - \, \int_{0}^{f_1(s_0, q^2)} \, {d \alpha_g \over \alpha_g } \,\,
{{\rm exp} \left ( - {s_0 / M^2 } \right ) \over 2 \, (m_b^2-q^2) (s_0-q^2)} \,  \,
 \rho_{V(A), 3}^{\rm 3P}
\left (s_0, q^2, \alpha_q=f_1(s_0, q^2)-\alpha_g, \alpha_{\bar q}, \alpha_g, v=1 \right ) \nonumber \\
&& \hspace{0.4 cm}  + \, \int_{0}^{1 - f_1(s_0, q^2)} \, {d \alpha_g \over \alpha_g } \,\,
{{\rm exp} \left ( - {s_0 / M^2 } \right ) \over 2 \, (m_b^2-q^2) (s_0-q^2)} \,  \,
 \rho_{V(A), 3}^{\rm 3P}
\left (s_0, q^2, \alpha_q=1-f_1(s_0, q^2), \alpha_{\bar q}, \alpha_g, v=0 \right ) \nonumber \\
&& \hspace{0.4 cm} + \, \int_0^{f_1(s_0, q^2)} \, d \alpha_q \, \int_{f_2(\alpha_q, s_0, q^2)}^{1-\alpha_q}
\, {d \alpha_g \over \alpha_g^2} \, {\theta(1- \alpha_q - f_2(\alpha_q, s_0, q^2)) \over (s_0-q^2)^2} \,
{\rm exp} \left ( - {s_0 \over  M^2 } \right )  \nonumber \\
&& \hspace{0.8 cm} \times \, {d \over d v} \, \left [ {1 \over 2 \, (\alpha_q + v \, \alpha_g)} \,
\rho_{V(A), 3}^{\rm 3P}
\left (s_0, q^2, \alpha_q, \alpha_{\bar q}, \alpha_g, v \right )  \right ]
\bigg |_{v=f_2(\alpha_q, s_0, q^2)/\alpha_g} \nonumber \\
&&  \hspace{0.4 cm}  + \,  \int_0^{f_1(s_0, q^2)} \, d \alpha_q  \,
\, \int_{f_2(\alpha_q, s_0, q^2)}^{1-\alpha_q}
\, {d \alpha_g \over \alpha_g} \, { s_0 -q^2 \over 2 \, (m_b^2-q^2)} \,
\theta(1- \alpha_q-f_2(\alpha_q, s_0, q^2))  \nonumber \\
&& \hspace{0.8 cm} \times \, {d \over d s_0} \,
\left [  {\rm exp} \left ( - {s_0 \over  M^2 } \right )  \, \rho_{V(A), 3}^{\rm 3P}
\left (s_0, q^2, \alpha_q, \alpha_{\bar q}, \alpha_g, v \right )  \right ]
\bigg |_{v=f_2(\alpha_q, s_0, q^2)/\alpha_g}  \nonumber \\
&& \hspace{0.4 cm} - \, \int_{m_b^2}^{s_0} \, ds  \,
\int_0^{f_1(s, q^2)} \, d \alpha_q  \,
\, \int_{f_2(\alpha_q, s, q^2)}^{1-\alpha_q}
\, {d \alpha_g \over \alpha_g} \, { s -q^2 \over 2 \, (m_b^2-q^2)} \,
\theta(1- \alpha_q-f_2(\alpha_q, s, q^2))  \nonumber \\
&& \hspace{0.8 cm} \times \, {d^2 \over d s^2} \,
\left [  {\rm exp} \left ( - {s \over  M^2 } \right )  \, \rho_{V(A), 3}^{\rm 3P}
\left (s, q^2, \alpha_q, \alpha_{\bar q}, \alpha_g, v \right )  \right ]
\bigg |_{v=f_2(\alpha_q, s, q^2)/\alpha_g} \bigg \}  \,,
\label{3-particle sum rules at LO}
\end{eqnarray}
where for brevity we have introduced the auxiliary function $f_2(\alpha_q, s, q^2)$ defined by
\begin{eqnarray}
f_2(\alpha_q, s, q^2) = {m_b^2 -q^2 \over s -q^2}  - \alpha_q \,.
\end{eqnarray}
In accordance with the power counting scheme for the threshold parameter and the
end-point behaviours of the three-particle photon DAs entering the sum rules
(\ref{3-particle sum rules at LO}), we can deduce the heavy-quark scaling of the
three-particle hadronic photon corrections
\begin{eqnarray}
F_{V, \, \rm photon}^{\rm 3P, \, LL}(n \cdot p) \sim
F_{A, \, \rm photon}^{\rm 3P, \, LL}(n \cdot p) \sim \left ({\Lambda \over m_b} \right )^{5/2} \,,
\end{eqnarray}
which is suppressed by one factor of $\Lambda/m_b$ compared with the higher-twist two-particle
contributions to the $B \to \gamma \ell \nu$ form factors
at tree level as presented in (\ref{LO sum rule for the 2PHT}).
It remains interesting to verify whether the NLO QCD corrections to the three-particle hadronic
photon contributions can give rise to a dynamically enhancement to remove the power-suppression
mechanism of the LO contributions (see \cite{Wang:2015vgv,Ball:2003bf} for a discussion
in the context of the NLO sum rules for the $B \to \pi$ form factors) and we will leave
explicit QCD calculations of the yet higher-order corrections  for future work.

Collecting the different pieces together, the resulting expressions for the $B \to \gamma \ell \nu$
form factors including the subleading-power contributions from the tree-level
$ b \, \bar u \to \gamma  \, W^{\ast}$ amplitude in QCD and from the hadronic photon corrections
can be written as
\begin{eqnarray}
F_V(n \cdot p) &=&  F_{V, \, \rm LP}(n \cdot p) + F_{V, \, \rm NLP}^{\rm LC}(n \cdot p)
+ F_{V, \, \rm photon}^{\rm 2PLT}(n \cdot p)
+ F_{V, \, \rm photon}^{\rm 2PHT, \, LL}(n \cdot p)
+ F_{V, \, \rm photon}^{\rm 3P, \, LL}(n \cdot p) \,, \nonumber \\
F_A(n \cdot p) &=& F_{A, \, \rm LP}(n \cdot p) + F_{A, \, \rm NLP}^{\rm LC}(n \cdot p)
+ F_{A, \, \rm photon}^{\rm 2PLT}(n \cdot p)
+ F_{A, \, \rm photon}^{\rm 2PHT, \, LL}(n \cdot p)
+ F_{A, \, \rm photon}^{\rm 3P, \, LL}(n \cdot p) \nonumber \\
&& + \,  {Q_{\ell} \, f_B \over v \cdot p} \,,
\label {final results of form factors}
\end{eqnarray}
where the last term proportional to the electric charge of the lepton
comes from the redefinition of the axial-vector form factor as discussed
in Section \ref{sect:Overview}. The detailed expressions of the individual terms
displayed on the right-hand side of (\ref{final results of form factors}) are
given by (\ref{resummation improved leading-power factorization formula}),
(\ref{local subleading power corrections}), (\ref{NLL sum rule at twist-2}),
(\ref{LO sum rule for the 2PHT}) and (\ref{3-particle sum rules at LO}),
respectively. We mention in passing that the LCSR calculations of the hadronic
photon corrections to the $B \to \gamma \ell \nu$ decay form factors presented here
suffer from the systematic uncertainty due to the parton-hadron duality ansatz
in the $B$-meson channel and future development of the subleading-power contributions
to the radiative leptonic $B$-meson decays in the framework of SCET including a proper
treatment of the rapidity divergences  appearing in the QCD factorization formulae
will be in demand for a model-independent QCD analysis.

\section{Numerical analysis}
\label{sect:numerical-analysis}

We are now ready to explore the phenomenological implications of the hadronic photon corrections
to the $B \to \gamma \ell \nu$ amplitude computed from the LCSR approach. To this end, we will proceed
by specifying the nonperturbative  models of the two-particle and three-particle photon DAs,
the first inverse moment $\lambda_B(\mu)$ and the logarithmic moments $\sigma_1(\mu)$ and $\sigma_2(\mu)$
of the leading-twist $B$-meson DA, and by determining the Borel mass and the hadronic threshold parameter
entering the sum rules for the subleading-power resolved photon contributions.
Having at our disposal the theory predictions for the form factors $F_V$ and $F_A$, we will further explore
the opportunity of constraining the inverse moment $\lambda_B(\mu)$ taking advantage of the improved measurements
at the Belle II  experiment in the near future.

\subsection{Theory inputs}

In analogy to the leading-twist photon DA, we employ the conformal expansion for the twist-three
DAs defined by the chiral-even light-cone matrix elements
\begin{eqnarray}
\psi^{(v)}(z, \mu) &=& 5 \, \left ( 3\, \xi^2 -1  \right )
+ {3 \over 64} \, \left ( 15 \, \omega_{\gamma}^{V}(\mu) - 5\, \omega_{\gamma}^{A}(\mu)  \right ) \,
\left ( 3 - 30 \, \xi^2 +  35 \,  \xi^4 \right ) \,,  \nonumber \\
\psi^{(a)}(z, \mu) &=&  {5 \over 2} \, \left ( 1 -  \xi^2 \right ) \, (5 \, \xi^2 -1 )
\left ( 1 + {9 \over 16} \, \omega_{\gamma}^{V}(\mu) - {3 \over 16} \, \omega_{\gamma}^{A}(\mu)  \right ) \,, \nonumber \\
V(\alpha_i, \mu) &=&  540 \,  \omega_{\gamma}^{V}(\mu) \, (\alpha_{q} - \alpha_{\bar q}) \,
\alpha_{q} \, \alpha_{\bar q} \, \alpha_g^2 \,,  \nonumber \\
A(\alpha_i, \mu) &=& 360 \, \alpha_{q} \, \alpha_{\bar q} \, \alpha_g^2 \,
\left [  1 + {\omega_{\gamma}^{A}(\mu) \over 2} \, (7 \, \alpha_g - 3) \right ]   \,,
\end{eqnarray}
with $\xi = 2 \, z -1$, and for the chiral-odd twist-four DAs
\begin{eqnarray}
\mathbb{A}(z, \mu) &=& 40 \, z^2 \, \bar z^2  \left [3 \, \kappa(\mu) - \kappa^{+}(\mu) + 1 \right ]
+ 8 \, \left [ \zeta_2^{+} (\mu) - 3 \, \zeta_2(\mu) \right ] \,
\big [ z \, \bar z \, (2 + 13 \,z \, \bar z )  \nonumber \\
&& + \, 2\, z^3 \, (10 - 15\, z + 6 \, z^2) \, \ln z
+ 2 \, \bar z^3 \, (10 - 15 \,\bar  z + 6 \, \bar z^2)  \, \ln \bar z \big  ] \,, \nonumber \\
h_{\gamma}(z, \mu) &=& -10 \, \left (1 + 2 \, \kappa^{+}(\mu) \right) \,
C_{2}^{1/2}(2 \, z -1)  \,, \nonumber \\
S(\alpha_i, \mu) &=& 30 \, \alpha_g^2 \,
\bigg \{ \left (\kappa(\mu) + \kappa^{+}(\mu) \right ) \, (1-\alpha_g)
+ (\zeta_1 + \zeta_1^{+}) (1 -\alpha_g) (1 -2 \, \alpha_g) \nonumber \\
&& + \, \zeta_2(\mu) \,\left [ 3 \, (\alpha_{\bar q} - \alpha_q)^2
- \alpha_g \, (1- \alpha_g) \right ] \bigg \}  \,, \nonumber \\
\widetilde{S}(\alpha_i, \mu) &=& - 30 \, \alpha_g^2 \,
\bigg \{ \left (\kappa(\mu) - \kappa^{+}(\mu) \right ) \, (1-\alpha_g)
+ (\zeta_1 - \zeta_1^{+}) (1 -\alpha_g) (1 -2 \, \alpha_g) \nonumber \\
&& + \, \zeta_2(\mu) \,\left [ 3 \, (\alpha_{\bar q} - \alpha_q)^2
- \alpha_g \, (1- \alpha_g) \right ] \bigg \} \,, \nonumber \\
S_{\gamma}(\alpha_i, \mu) &=& 60 \, \alpha_g^2 \, (\alpha_q + \alpha_{\bar q} ) \,
\left [ 4 - 7 \, (\alpha_{\bar q} + \alpha_q )  \right ] \,, \nonumber \\
T_{1}(\alpha_i, \mu) &=& - 120 \, \left (3 \, \zeta_2(\mu) + \zeta_2^{+}(\mu) \right ) \,
\left ( \alpha_{\bar q} - \alpha_q \right )  \, \alpha_{\bar q} \, \alpha_q \, \alpha_g \,, \nonumber \\
T_{2}(\alpha_i, \mu) &=& 30 \, \alpha_g^2 \, (\alpha_{\bar q} - \alpha_q )
  \left [  \left (\kappa(\mu) - \kappa^{+}(\mu) \right )
+ \left (\zeta_1(\mu) - \zeta_1^{+}(\mu) \right )(1-2 \, \alpha_g)
+ \zeta_2(\mu) \, (3 -4 \, \alpha_g)\right ] \,, \nonumber \\
T_{3}(\alpha_i, \mu) &=& -120 \, \left (3 \, \zeta_2(\mu) - \zeta_2^{+}(\mu) \right ) \,
(\alpha_{\bar q} - \alpha_q ) \,\alpha_{\bar q} \, \alpha_q \, \alpha_g \,, \nonumber \\
T_{4}(\alpha_i, \mu) &=& 30 \, \alpha_g^2  \, (\alpha_{\bar q} -\alpha_q)\,
\left [  \left (\kappa(\mu) + \kappa^{+}(\mu) \right )
+ \left (\zeta_1(\mu) + \zeta_1^{+}(\mu) \right )(1-2 \, \alpha_g)
+ \zeta_2(\mu) \, (3 - 4 \, \alpha_g)\right ]  \,, \nonumber \\
T_{4}^{\gamma}(\alpha_i, \mu) &=& 60 \, \alpha_g^2 \, (\alpha_q - \alpha_{\bar q} ) \,
\left [ 4 - 7 \, (\alpha_{\bar q} + \alpha_q )  \right ]  \,.
\end{eqnarray}
Here, we have truncated the conformal expansion of the photon light-cone DAs up to the
next-to-leading conformal spin (i.e., ``P"-wave).
The renormalization-scale dependence of the twist-three parameters can be written as
\begin{eqnarray}
&& f_{3 \gamma}(\mu)= \left (  {\alpha_s (\mu) \over \alpha_s (\mu_0)} \right )^{\gamma_f  / \beta_0} \,
 f_{3 \gamma}(\mu_0)\,,  \qquad
\gamma_f = - {C_F \over 3} + 3\, C_A \,, \qquad
\beta_0 = 11 - {2 \, n_f  \over 3}\,,
\nonumber \\
&& \left(
\begin{array}{c}
\omega^{V}_\gamma(\mu) - \omega^{A}_\gamma(\mu) \\
\omega^{V}_\gamma(\mu) + \omega^{A}_\gamma(\mu)
\end{array}
\right)
= \left (  {\alpha_s (\mu) \over \alpha_s (\mu_0)} \right )^{\Gamma_{\omega}  / \beta_0} \,
\left(
\begin{array}{c}
\omega^{V}_\gamma(\mu_0) - \omega^{A}_\gamma(\mu_0) \\
\omega^{V}_\gamma(\mu_0) + \omega^{A}_\gamma(\mu_0)
\end{array}
\right) \,,
\end{eqnarray}
where the anomalous dimension matrix $\Gamma_{\omega}$ is given by \cite{Ball:2002ps,Ball:1998sk}
\begin{eqnarray}
\Gamma_{\omega}=
\left(
\begin{array}{c}
3 \, C_F - {2 \over 3} \, C_A  \qquad   {2 \over 3} \, C_F - {2 \over 3} \, C_A   \\
{5 \over 3} \, C_F - {4 \over 3} \, C_A \qquad  {1 \over 2} \, C_F  + C_A
\end{array}
\right)\,.
\end{eqnarray}
Due to the Ferrara-Grillo-Parisi-Gatto theorem  \cite{Ferrara:1972xq}, the twist-four parameters corresponding to
the ``P"-wave conformal spin satisfies the following relations
\begin{eqnarray}
\zeta_1(\mu) + 11 \,\zeta_2(\mu) - 2 \, \zeta_2^{+}(\mu) = {7 \over 2} \,.
\end{eqnarray}
The scale evolution of the nonperturbative parameters at twist-four accuracy is given by
\begin{eqnarray}
\kappa^{+}(\mu)= \left (  {\alpha_s (\mu) \over \alpha_s (\mu_0)} \right )
^{\left ( \gamma^{+} - \gamma_{q \bar q} \right )  / \beta_0} \,
\kappa^{+}(\mu_0)\,, & \qquad &
\kappa(\mu)= \left (  {\alpha_s (\mu) \over \alpha_s (\mu_0)} \right )
^{\left ( \gamma^{-} - \gamma_{q \bar q} \right )  / \beta_0} \,
\kappa(\mu_0) \,, \nonumber \\
\zeta_1(\mu)= \left (  {\alpha_s (\mu) \over \alpha_s (\mu_0)} \right )
^{\left ( \gamma_{Q^{(1)}} - \gamma_{q \bar q} \right )  / \beta_0} \,
\zeta_1(\mu_0)\,,  & \qquad &
\zeta_1^{+}(\mu)= \left (  {\alpha_s (\mu) \over \alpha_s (\mu_0)} \right )
^{\left ( \gamma_{Q^{(5)}} - \gamma_{q \bar q} \right )  / \beta_0} \,
\zeta_1^{+}(\mu_0)\,, \nonumber \\
\zeta_2^{+}(\mu)= \left (  {\alpha_s (\mu) \over \alpha_s (\mu_0)} \right )
^{\left ( \gamma_{Q^{(3)}} - \gamma_{q \bar q} \right )  / \beta_0} \,
\zeta_2^{+}(\mu_0)\,,
\label{evolution of twist-four parameters}
\end{eqnarray}
where the anomalous dimensions of these twist-four parameters at one loop are given by \cite{Ball:2002ps}
\begin{eqnarray}
\gamma^{+}= 3 \, C_A - {5 \over 3} \, C_F \,, & \qquad &
\gamma^{-}= 4 \, C_A - 3 \, C_F \,, \nonumber \\
r_{q \bar q} =-3 \, C_F \,, & \qquad &
\gamma_{Q^{(1)}} = {11 \over 2} \, C_A - 3 \, C_F \,,
\nonumber \\
\gamma_{Q^{(3)}} = {13 \over 3} \, C_F \,, & \qquad &
\gamma_{Q^{(5)}} = 5 \, C_A - {8 \over 3} \, C_F \,.
\end{eqnarray}
Numerical values of the input parameters entering the photon DAs up to twist-four
are collected in Table \ref{tab of parameters for photon DAs},
where we have assigned 100 \% uncertainties for the estimates of
the twist-four parameters from QCD sum rules \cite{Balitsky:1989ry}.
The second Gegenbauer moment of the leading-twist photon DA will be further taken as
$a_2(\mu_0)=0.07 \pm 0.07$  as obtained in \cite{Ball:2002ps}.
The magnetic susceptibility of the quark condensate $\chi (1 \, {\rm GeV}) = (3.15 \pm 0.3) \, {\rm GeV}^{-2}$
computed  from the QCD sum rule approach including the ${\cal O}(\alpha_s)$ corrections \cite{Ball:2002ps}
and the quark condensate density $\langle \bar q q \rangle (1 \, {\rm GeV})=-(246^{+28}_{-19} \, {\rm MeV})^3$
determined by the GMOR relation  \cite{Duplancic:2008ix} will be also employed for the numerical estimates
in the following.

\begin{table}[t]
\begin{center}
\begin{tabular}{|c|c|c|c|c|c|c|c|}
  \hline
  \hline
&&&&&&& \\[-3.5mm]
$f_{3 \gamma}(\mu_0) \, ({\rm GeV^2})$ & $\omega_{\gamma}^{V}(\mu_0) $ & $\omega_{\gamma}^{A}(\mu_0)$ & $\kappa(\mu_0)$
& $\kappa^{+}(\mu_0)$ & $\zeta_{1}(\mu_0)$ &
$\zeta_{1}^{+}(\mu_0)$ &
$\zeta_{2}^{+}(\mu_0)$  \\
    &&&&&&& \\[-1mm]
  \hline
    &&&&&&& \\[-1mm]
  $- (4 \pm 2) \, \times 10^{-3}$ &$3.8 \pm 1.8$ & $-2.1 \pm 1.0$ & $0.2 \pm 0.2$ &$0$&
$0.4 \pm 0.4$ & $0$  & $0$ \\
&&&&&&& \\
\hline
\hline
\end{tabular}
\end{center}
\caption{Numerical values of the nonperturbative parameters entering the photon DAs
at the renormalization scale $\mu_0= 1.0 \, {\rm GeV}$.}
\label{tab of parameters for photon DAs}
\end{table}

The key quantity entering the leading-power factorization formula of the $B \to \gamma \ell \nu$
form factors is the first inverse moment of the $B$-meson DA $\lambda_B(\mu)$, whose determination
has been discussed extensively in the context of exclusive $B$-meson decays with distinct QCD approaches
(see \cite{Wang:2015vgv,Wang:2017jow} for more discussions). To illustrate the phenomenological consequences
of the subleading-power corrections from the hadronic photon contributions we will take the interval
$\lambda_B(1 \, {\rm GeV})=354^{+38}_{-30} \, {\rm MeV}$ implied by the LCSR calculations of the semileptonic
$B \to \pi$ form factors with $B$-meson DAs on the light-cone \cite{Wang:2015vgv}. The renormalization-scale
dependence of $\lambda_B(\mu)$ at one loop can be determined from the evolution equation of
 $\phi_B^{+}(\omega ,\mu)$ \cite{Lange:2003ff}
\begin{eqnarray}
{\lambda_B(\mu) \over \lambda_B(\mu_0)} =
1 + {\alpha_s(\mu_0) \, C_F \over 4 \, \pi} \, \ln {\mu \over \mu_0} \,
\left [ 2 - 2 \, \ln {\mu \over \mu_0} - 4 \, \sigma_1(\mu_0) \right ] + {\cal O}(\alpha_s^2) \,,
\end{eqnarray}
where the inverse-logarithmic moments $\sigma_n(\mu_0)$ are defined as \cite{Beneke:2011nf}
\begin{eqnarray}
\sigma_n(\mu_0) = \lambda_B(\mu) \, \int_0^{\infty} \, {d \omega \over \omega} \, \ln^n \left ({\mu_0 \over \omega} \right ) \,
\phi_B^{+}(\omega, \mu_0) \,.
\end{eqnarray}
Numerically we will employ  $\sigma_1(1 {\rm GeV})=1.5 \pm 1$ consistent with the NLO QCD sum rule calculation
\cite{Braun:2003wx} and $\sigma_2(1 {\rm GeV})= 3 \pm 2$ from \cite{Beneke:2011nf}.
Furthermore, the static $B$-meson decay constant $\tilde{f}_B(\mu)$ will be expressed in terms of the QCD decay constant $f_B$
\begin{eqnarray}
\tilde{f}_B(\mu)=f_B \, \left \{ 1 + {\alpha_s(\mu) \, C_F \over 4 \, \pi}  \,
\left [ 3 \, \ln {m_B \over \mu}  - 2 \right ]  \right \}^{-1} \,,
\end{eqnarray}
and the determination $f_B= (192.0 \pm 4.3) \, {\rm MeV}$ from the FLAG Working Group \cite{Aoki:2016frl}
will be taken in the numerical analysis.

Following the discussions presented in \cite{Beneke:2011nf,Wang:2015vgv}, the hard scales $\mu_{h1}$ and $\mu_{h2}$ entering
the leading-power factorization formula  will be chosen as
$\mu_{h1}=\mu_{h2} \in  [m_b/2, \,\,\, 2 \, m_b]$ around the default value $m_b$ and the factorization scale
in (\ref{resummation improved leading-power factorization formula})  will be
varied in the interval $1\, {\rm GeV} \leq \mu \leq 2\, {\rm GeV}$ with the central value $\mu = 1.5 \, {\rm GeV}$.
In contrast, the factorization scale entering the LCSR for the hadronic photon corrections will be taken as
$\mu \in [m_b/2, \,\,\, 2 \, m_b]$ around the default choice $m_b$.
In addition, we adopt the numerical values of the bottom quark mass
$\overline{m_b}(\overline{m_b})=4.193^{+0.022}_{-0.033} \, {\rm GeV}$ \cite{Beneke:2014pta}
in the ${\rm \overline{MS}}$ scheme from non-relativistic sum rules.
Finally, we turn to determine the Borel mass $M^2$ and the threshold parameter $s_0$
 in the LCSR for the hadronic photon contributions. Applying the standard strategies presented
 in \cite{Wang:2015vgv} (see also \cite{Colangelo:2000dp} for a review) gives rise to following intervals
\begin{eqnarray}
s_0 = (37.5 \pm 2.5) \, {\rm GeV^2} \,, \qquad
M^2 = (18.0 \pm 3.0) \, {\rm GeV^2}  \,,
\end{eqnarray}
which is consistent with the determinations from the LCSR of the $B \to \pi$ form factors \cite{Khodjamirian:2011ub}.

\subsection{Predictions for the $B \to \gamma \ell \nu$ form factors}

\begin{figure}
\begin{center}
\includegraphics[width=0.58 \columnwidth]{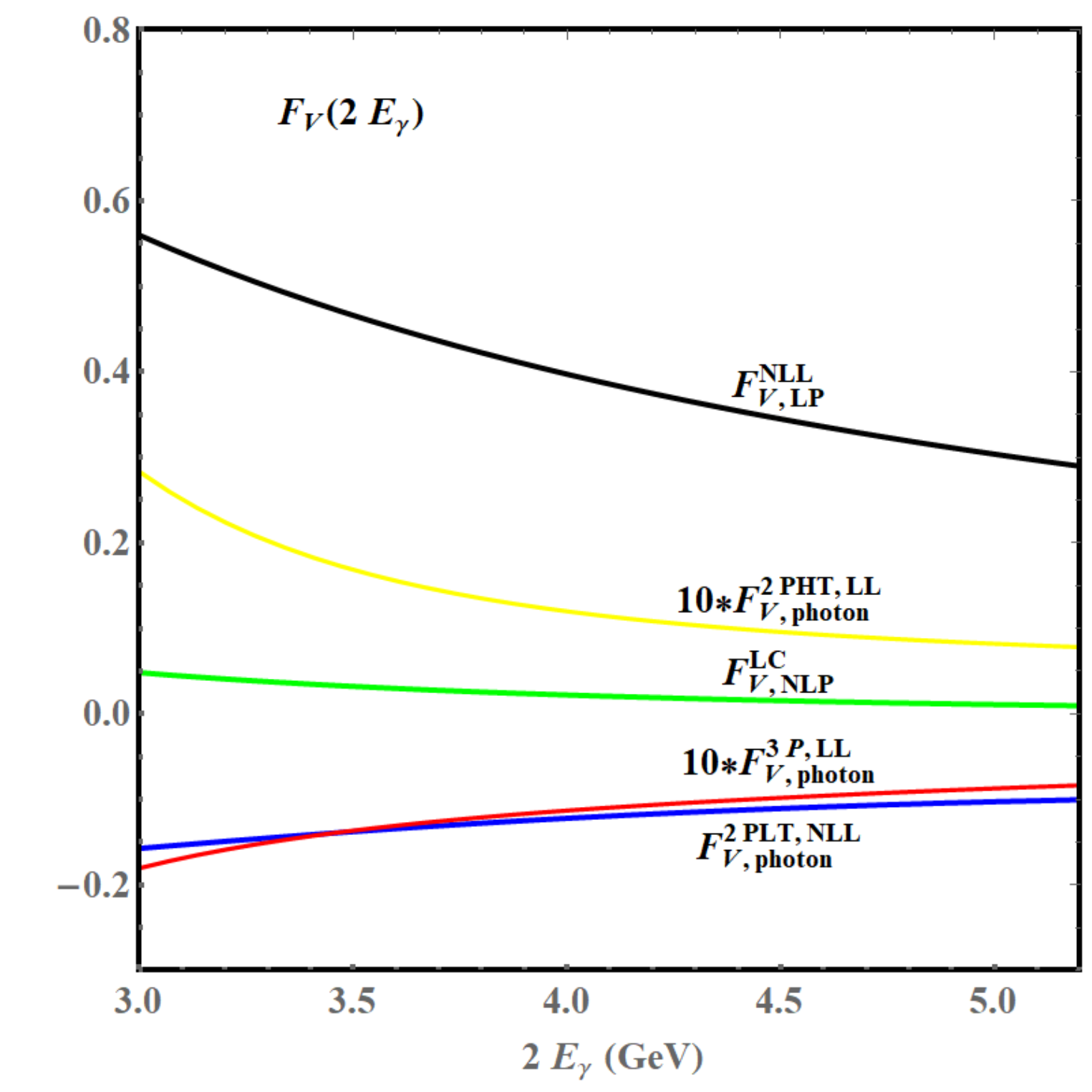} \\
\vspace*{0.1cm}
\caption{The photon-energy dependence of different terms contributing to the vector
$B \to \gamma$ form factor $F_V(2 \, E_{\gamma})$ as displayed in (\ref{final results of form factors})
with the central values of theory inputs.
The individual contributions correspond to the leading-power contribution at NLL computed from the
QCD factorization approach ($F_{V, \, {\rm LP}}^{\rm NLL}$, black), the subleading-power local contribution
at LO ($F_{V, \, {\rm NLP}}^{\rm LC}$, green), the two-particle leading-twist hadronic photon correction
at NLL ($F_{V, \, {\rm photon}}^{\rm 2PLT, \, NLL}$,  blue),
the two-particle higher-twist hadronic photon correction at leading-logarithmic (LL) accuracy
($F_{V, \, {\rm photon}}^{\rm 2PHT, \, LL}$,  yellow),
the three-particle leading-twist hadronic photon correction
at LL ($F_{V, \, {\rm photon}}^{\rm 3P, \, LL}$, red).}
\label{fig: photon-energy dependence of the vector form factor}
\end{center}
\end{figure}

We are now in a position to explore the phenomenological significance of the hadronic photon corrections
to the $B \to \gamma \ell \nu$ form factors. To develop a better understanding of the heavy quark expansion
for the bottom sector, we plot the photon-energy dependence of the leading-power contribution,
the subleading-power local correction and the subleading-power two-particle and three-particle
hadronic photon effects in figure \ref{fig: photon-energy dependence of the vector form factor}.
It is apparent that the twist-two hadronic photon contribution at NLL can generate sizeable destructive interference
with the leading-power ``direct photon" contribution: approximately ${\cal O}(30 \%)$ for
$n \cdot p \in [3 \, {\rm GeV} \,, m_B]$ with $\lambda_B(\mu_0)=354 \, {\rm MeV}$.
However, both the two-particle higher-twist and the three-particle hadronic photon contributions turn out to
be numerically insignificant  at tree level and will only shift the leading-power prediction by an amount of
${\cal O} \, (3 \sim 5) \%$ for $n \cdot p \in [3 \, {\rm GeV} \,, m_B]$.
Furthermore, the subleading-power local contribution $F_{V, \, {\rm NLP}}^{\rm LC}$ at tree level displayed in
(\ref{local subleading power corrections}) will give rise to ${\cal O} (3 \%)$ correction at
$n \cdot p = m_B$ and ${\cal O} (10 \%)$ correction at $n \cdot p = 3 \, {\rm GeV}$.
On account of the observed pattern for the separate terms contributing to the
$B \to \gamma $ form factors numerically, we are led to conclude that the power suppressed contributions
to the radiative leptonic $B$-meson decay are dominated by the leading-twist hadronic photon correction
with the default theory inputs.

\begin{figure}
\begin{center}
\includegraphics[width=0.45 \columnwidth]{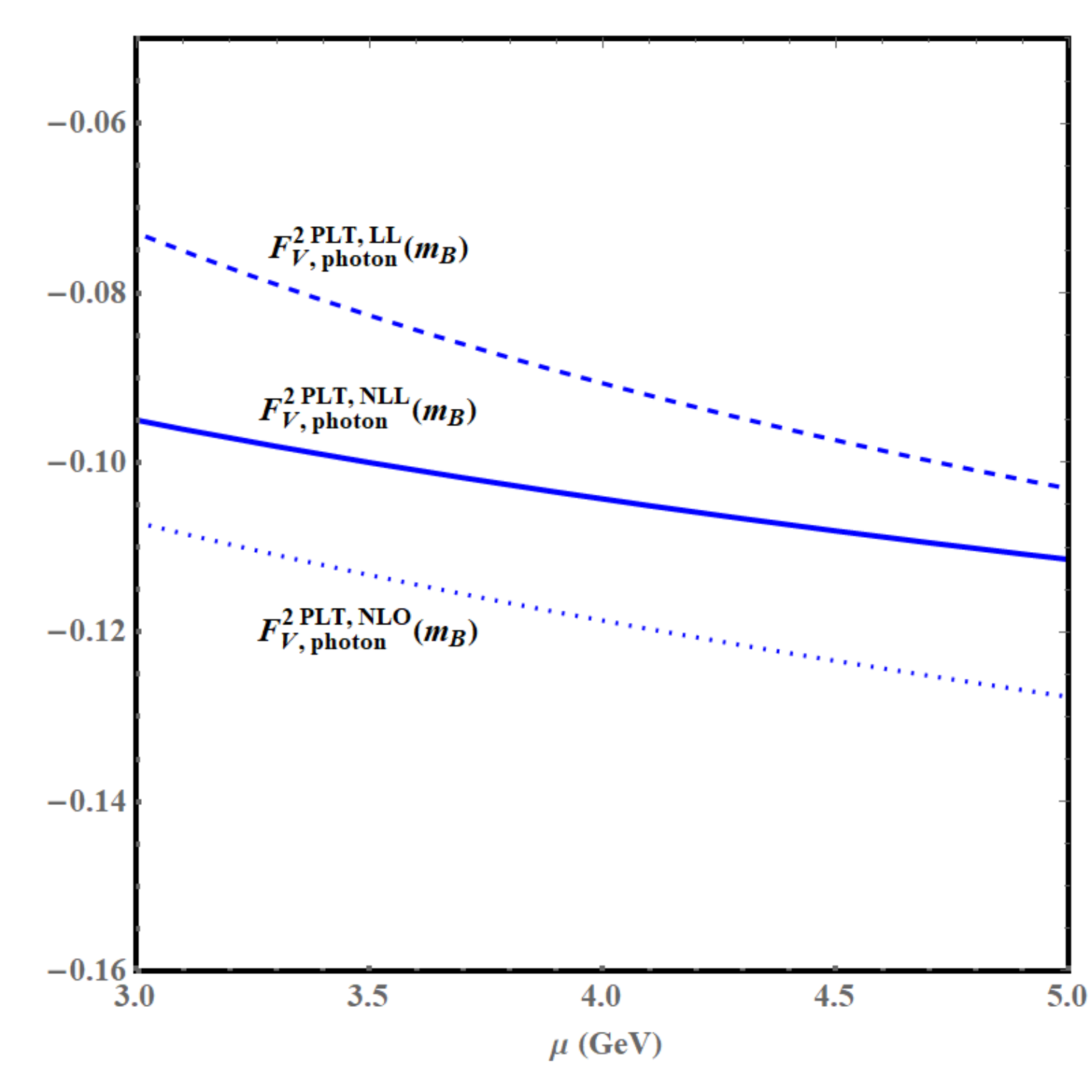}
\hspace{0.4 cm}
\includegraphics[width=0.44 \columnwidth]{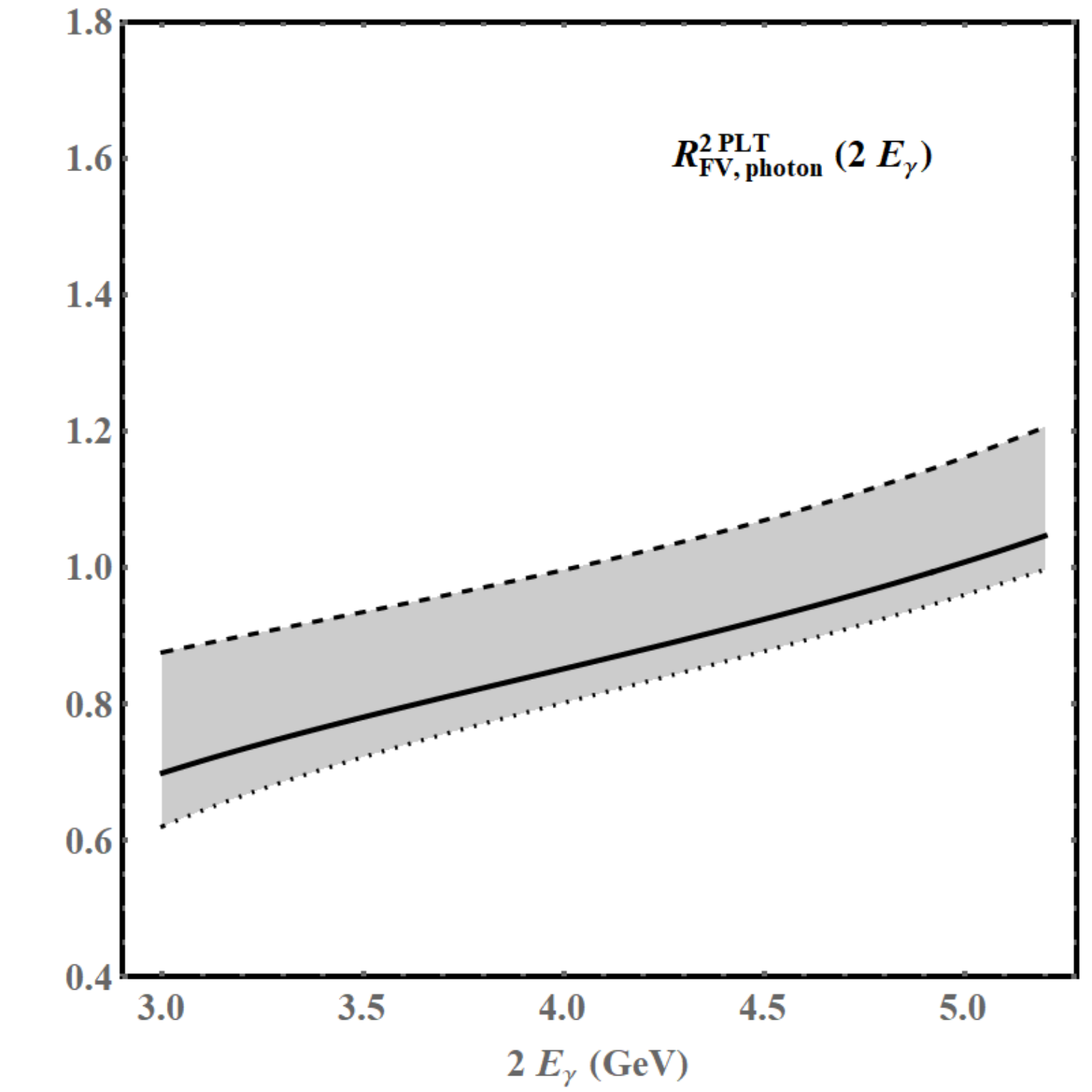}\\
\vspace*{0.1cm}
\caption{{\it Left}:  The renormalization scale dependence of the twist-two hadronic
photon correction to the vector $B \to \gamma$ form factor $F_{V}(m_B)$ at LL (dashed),
NLO (dotted), and  NLL (solid) accuracy, respectively. {\it Right}: The photon-energy
dependence of the ratio $R_{FV, \, {\rm photon}}^{\rm 2PLT}(n \cdot p) \equiv
F_{V, \, {\rm photon}}^{\rm 2PLT, \, NLL}(n \cdot p)/F_{V, \, {\rm photon}}^{\rm 2PLT, \, LL}(n \cdot p)$
with the uncertainties  from the variations of the renormalization scale $\mu$.  }
\label{fig: resummation and NLO QCD effects}
\end{center}
\end{figure}

We further turn to investigate the numerical impact of the perturbative correction at NLO and the
QCD resummation of the parametrically large logarithms of $m_b^2/\Lambda^2$ for the leading-twist
hadronic photon contribution computed from the LCSR technique.
It is evident that from figure \ref{fig: resummation and NLO QCD effects} that the NLO QCD correction
can  decrease the tree-level prediction of the twist-two hadronic photon contribution by an amount of
${\cal O} \, (20 \sim 40) \%$ for the factorization scale varied in the interval $[3.0, \, 5.0] \, {\rm GeV}$
 and the NLL resummation effect can yield ${\cal O} \, (10 \, \%)$ enhancement
to the NLO QCD results within the same range of $\mu$. Hence,  the dominant radiative correction
to the leading-twist hadronic photon contribution originates from the NLO QCD correction to the hard matching coefficient
entering the factorization formula (\ref{NLO factorization formula at LT}) rather than from
resummation of the large logarithms $m_b^2/\Lambda^2$.
However, the renormalization scale dependence of the resummation improved theory predictions
in the allowed region indeed becomes weaker compared  with the NLO calculation.
We further plot the photon-energy dependence of the ratio
$R_{FV, \, {\rm photon}}^{\rm 2PLT}(n \cdot p) \equiv
F_{V, \, {\rm photon}}^{\rm 2PLT, \, NLL}(n \cdot p)/F_{V, \, {\rm photon}}^{\rm 2PLT, \, LL}(n \cdot p)$
characterizing the perturbative QCD corrections at NLL in figure \ref{fig: resummation and NLO QCD effects},
where the theory uncertainties due to the variations of the renormalization scale $\mu$ are also displayed.

\begin{figure}
\begin{center}
\includegraphics[width=0.45 \columnwidth]{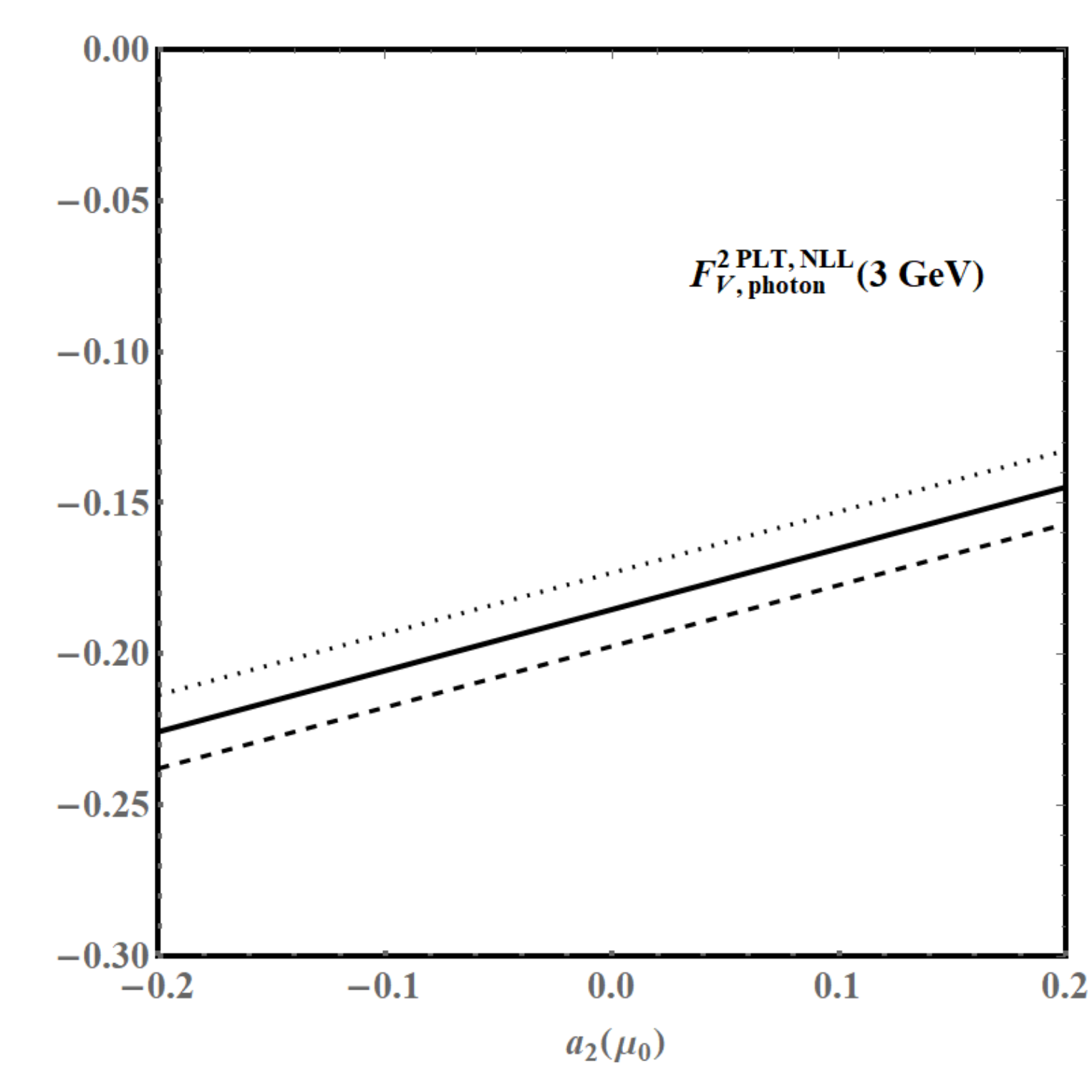}
\hspace{0.4 cm}
\includegraphics[width=0.45 \columnwidth]{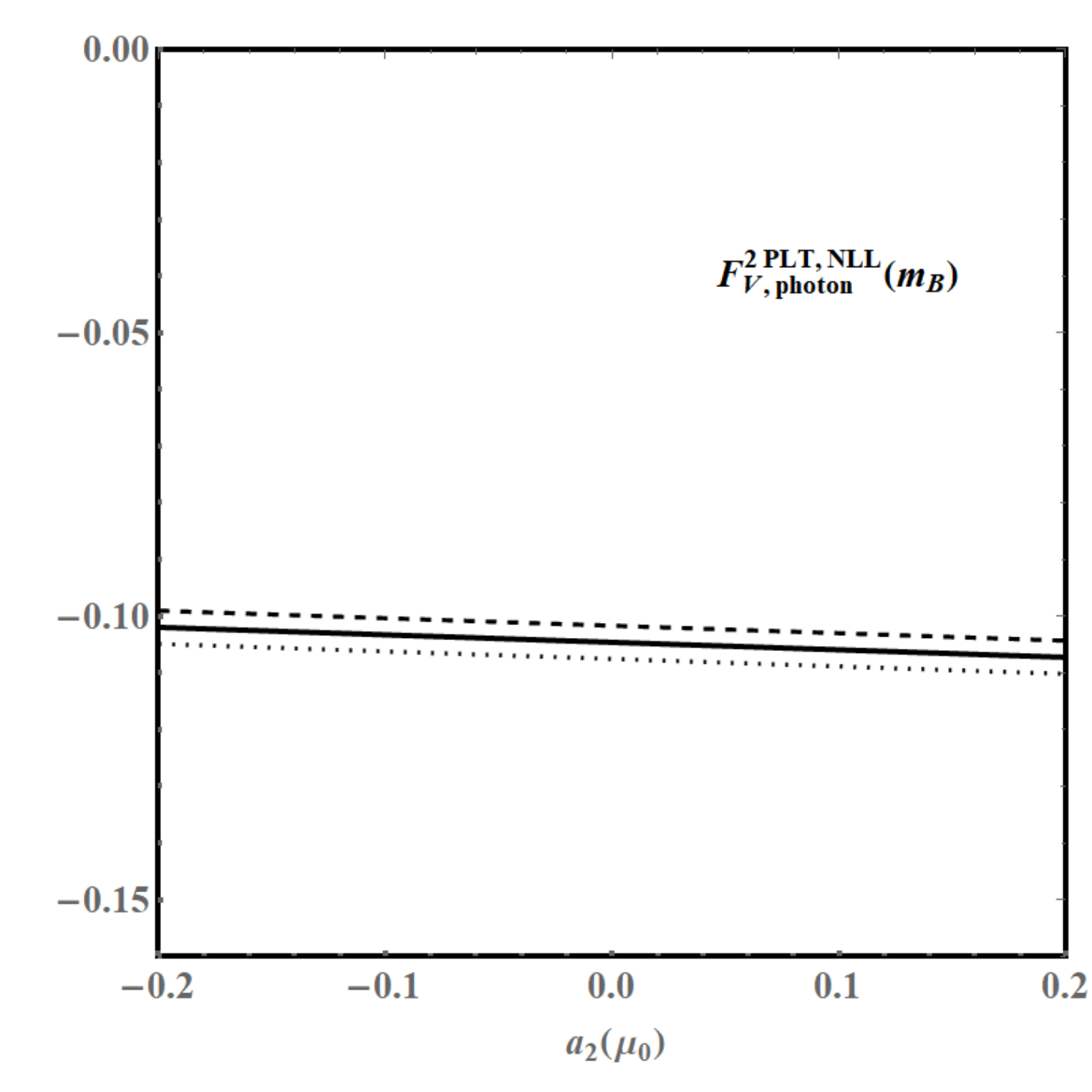}\\
\vspace*{0.1cm}
\caption{Dependence of the leading-twist hadronic photon correction
at $n\cdot p = 3 \, {\rm GeV}$ (left panel) and $n\cdot p = m_B$ (right panel)
on the second Gegenbauer moment of the photon light-cone DA $a_2(\mu_0)$ with
different values of the fourth Gegenbauer moment: $a_4(\mu_0)=0.2$ (dashed),
$a_4(\mu_0)=0$ (solid) and $a_4(\mu_0)=-0.2$ (dotted).}
\label{fig: a2 dependence of B to gamma form factors}
\end{center}
\end{figure}

Taking into account the fact that the QCD sum rule calculation of the second Gegenbauer moment of
the twist-two photon DA $a_2(\mu_0)$ suffers from the large theory uncertainties due to the strong sensitivity
to the input parameters \cite{Ball:2002ps}, we plot the leading-twist hadronic photon correction
to the vector form factor $F_V(n \cdot p)$ in a wide range of $a_2(\mu_0)$ in figure
\ref{fig: a2 dependence of B to gamma form factors}.
One can readily observe that the variation of the  Gegenbauer moment $a_2(\mu_0) \in [-0.2, 0.2]$
can only give rise to a minor impact on the theory prediction of the
$B \to \gamma$ form factor $F_V(m_B)$ at maximal recoil numerically.
However, the ``P-wave" conformal spin contribution from the leading-twist photon DA will become
significant for the evaluation of the form factor $F_V(n \cdot p)$ with the decrease of the photon energy:
approximately ${\cal O}(35 \%)$ at $n \cdot p = 3 \, {\rm GeV}$.
To further understand the systematic uncertainty due to the truncation of the conformal expansion at ``P-wave",
we also display the theory predictions for the $B \to \gamma \ell \nu$ form factors including the  ``D-wave"
effect from the fourth Gegenbauer moment $a_4(\mu_0)$ in figure \ref{fig: a2 dependence of B to gamma form factors}.
It is apparent that the sensitivity of the leading-twist hadronic photon contribution on $a_4(\mu_0)$ is rather
weak numerically  for $n \cdot p \in [3 \, {\rm GeV}, m_B]$  in the ``reasonable" interval $-0.2 \leq a_4(\mu_0) \leq 0.2$.
In the light of such observation, the yet higher Gegenbauer moments of the twist-two photon DA are not expected to bring about
notable impact on the prediction of the subleading-power contribution to the $B \to \gamma \ell \nu$ form factors
induced by the photon light-cone DAs.

\begin{figure}
\begin{center}
\includegraphics[width=0.45 \columnwidth]{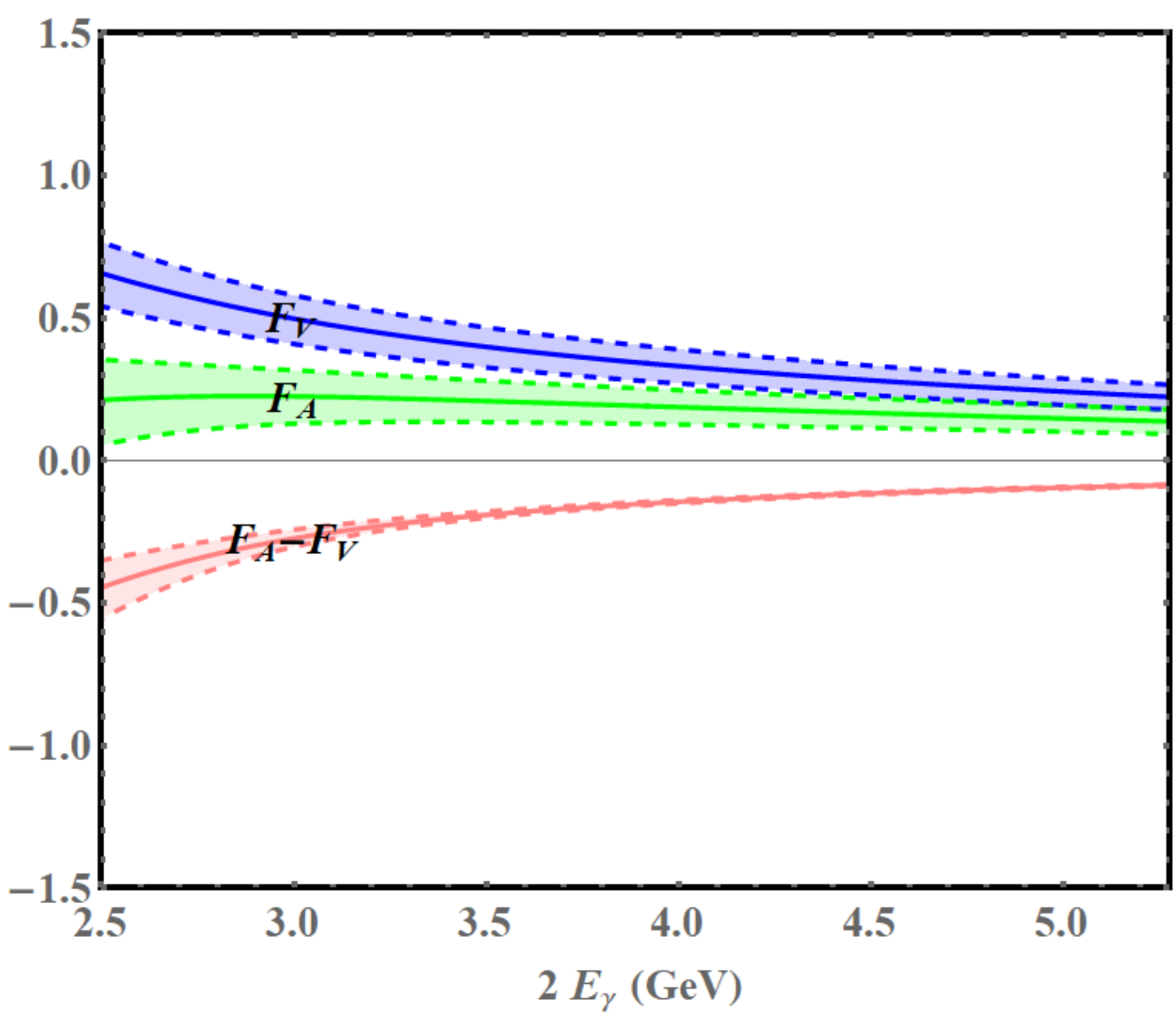}
\caption{The photon-energy dependence of the $B \to \gamma \ell \nu$ form factors
as well as their difference computed from (\ref{final results of form factors})
with the theory uncertainties from  variations of different input parameters added in quadrature.}
\label{fig: uncertainties of B to gamma form factors}
\end{center}
\end{figure}

\begin{figure}
\begin{center}
\includegraphics[width=0.45 \columnwidth]{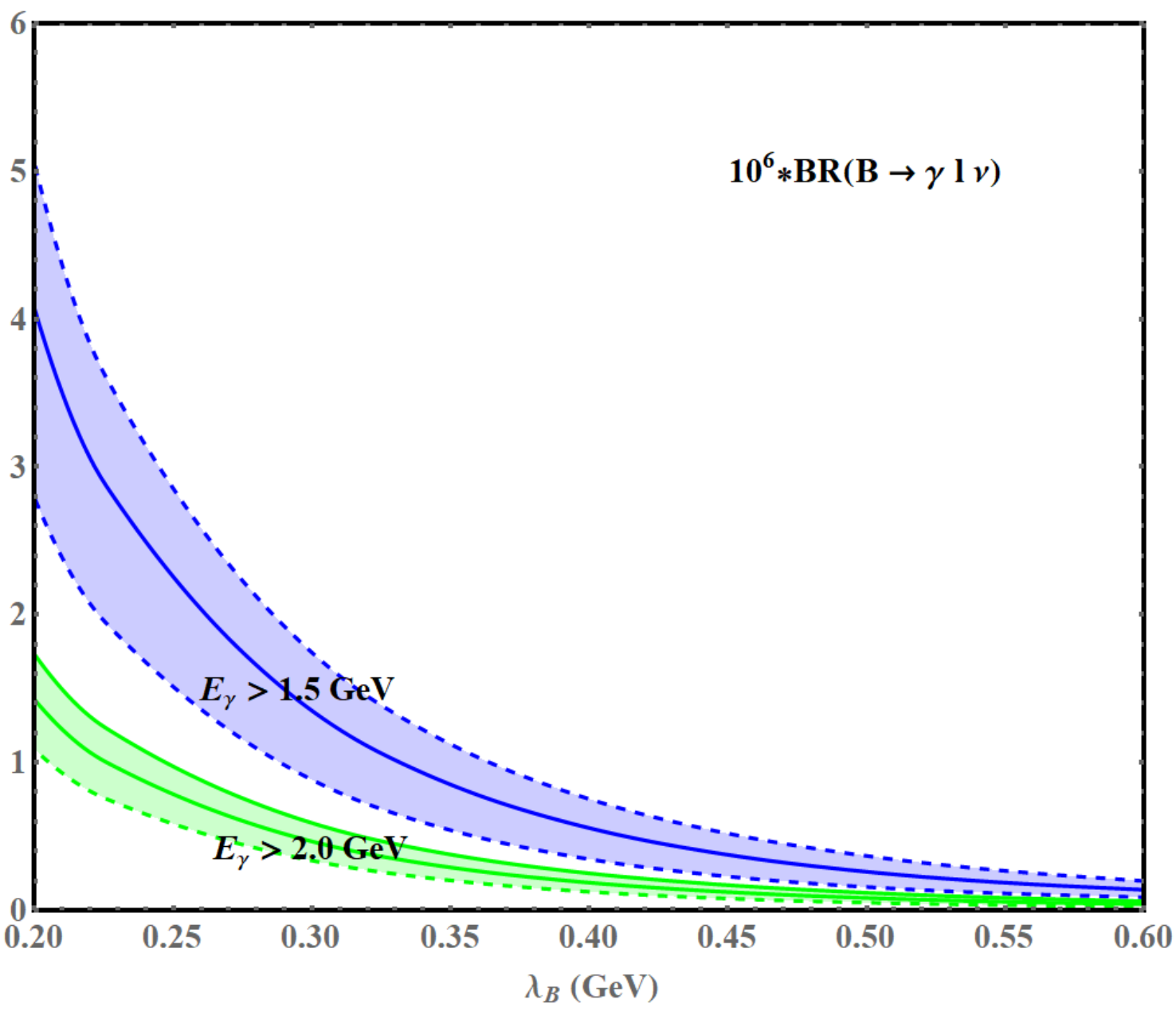}
\caption{Dependence of the partial branching fractions
${\cal BR}(B \to \gamma \ell \nu, \, E_{\gamma} \geq E_{\rm cut})$ on
the first inverse moment $\lambda_B(\mu_0)$ for $E_{\rm cut}=1.5 \, {\rm GeV}$
(blue band) and $E_{\rm cut}=2.0 \, {\rm GeV}$ (green band).}
\label{fig: predicted branching ratio}
\end{center}
\end{figure}

We present our final predictions for the $B \to \gamma \ell \nu$ form factors including the newly computed
two-particle and three-particle hadronic photon corrections with theory uncertainties  in
figure \ref{fig: uncertainties of B to gamma form factors}. The dominant theory uncertainties originate from
the first inverse moment $\lambda_B(\mu_0)$, the factorization scale $\mu$ entering the leading-power
``direct photon" contribution, and the second Gegenbauer moment $a_2(\mu_0)$ of the twist-two photon DA.
However, the symmetry breaking effect between the two $B \to \gamma$ form factors due to the subleading-power local contribution
and the higher-twist hadronic photon corrections suffers from much less uncertainty than the individual form factors at
$3 \, {\rm GeV} \leq  n \cdot p \leq m_B$. Having in our hands the theoretical predictions for the
$B \to \gamma \ell \nu$ form factors, we proceed to discuss the theory constraints on the inverse moment $\lambda_B(\mu_0)$
taking advantage of the future measurements on the (partially) integrated branching fractions with a photon-energy cut
to get rid of the soft photon radiation. It is straightforward to derive the differential decay width for $B \to \gamma \ell \nu$
in the rest frame of the $B$-meson (see also \cite{Beneke:2011nf,Wang:2016qii})
\begin{eqnarray}
{d \, \Gamma(B \to \gamma \ell \nu) \over d \, E_{\gamma}}
= {\alpha_{em} \, G_F^2 \, |V_{ub}|^2 \over 6\, \pi^2} \, m_B \,
E_{\gamma}^3 \, \left (  1- { 2 \, E_{\gamma} \over m_B} \right ) \,
\left [F_V^2( n \cdot p) + F_A^2( n \cdot p)   \right ] \,,
\end{eqnarray}
and the integrated branching fractions with the phase-space cut on the
photon energy read
\begin{eqnarray}
{\cal BR}(B \to \gamma \ell \nu, \, E_{\gamma} \geq E_{\rm cut})
= \tau_B \, \int_{E_{\rm cut}}^{m_B/2} \, d E_{\gamma} \,
{d \, \Gamma(B \to \gamma \ell \nu) \over d \, E_{\gamma}} \,,
\end{eqnarray}
where $\tau_B$ indicates the lifetime of the $B$-meson.
Our predictions for the partial branching fractions of
the radiative leptonic decay $B \to \gamma \ell \nu$ including the hadronic photon
corrections to the form factors are displayed in figure \ref{fig: predicted branching ratio}
with the variation of the inverse moment $\lambda_B(\mu_0)$ in the interval $[0.2, 0.6] \, {\rm GeV}$.
It can be observed that the integrated branching fractions ${\cal BR}(B \to \gamma \ell \nu, \, E_{\gamma} \geq E_{\rm cut})$
grow rapidly  with the decrease of the inverse moment due to the dependence of the two form factors on
$1 / \lambda_B(\mu_0)$ at leading-power in $\Lambda/m_b$.
Since the photon-energy cut $E_{\gamma} \geq 1 \, {\rm GeV}$ implemented in the Belle measurements \cite{Heller:2015vvm}
is not sufficiently large to perform perturbative QCD calculations of  the $B \to \gamma$ form factors,
we will not employ the  experimental bound ${\cal BR}(B \to \gamma \ell \nu, \, E_{\gamma} \geq E_{\rm cut}) < 3.5 \times 10^{-6}$
 with the full Belle data sample reported in \cite{Heller:2015vvm} for the determination of $\lambda_B(\mu_0)$ at the moment.
Instead, we prefer to explore the solid theory constraints on the first inverse moment by comparing our predictions of the
(partially) integrated branching fractions with the improved measurements  at the Belle II experiment,
with the tighter phase-space cut on the photon energy, thanks to the much higher designed
luminosity of the SuperKEKB accelerator.

\section{Conclusion}
\label{sect:Conclu}

We computed perturbative QCD corrections to the leading-twist hadronic photon contribution to the $B \to \gamma \ell \nu$
 form factors employing the LCSR method. QCD factorization for the vacuum-to-photon correlation function
 (\ref{def: correlation function}) has been demonstrated explicitly at one loop with the OPE technique
 and the NDR scheme of the Dirac matrix $\gamma_5$ including the evanescent SCET operator.
 The perturbative matching coefficient entering the NLO factorization formula (\ref{NLO factorization formula at LT})
 was obtained by applying the method of regions and the factorization-scale independence of the correlation function
 (\ref{def: correlation function})  was further verified at ${\cal O} (\alpha_s)$ with the evolution equations of
 the twist-two photon DA and the bottom-quark mass.
 Resummation of the parametrically large logarithms of ${\cal O} (\ln (m_b^2/\Lambda^2))$ was achieved at NLL accuracy
with the two-loop RG equation of the  light-ray tensor operator. Implementing the continuum subtraction with the aid of
the parton-hadron duality and the Borel transformation, the NLL resummation improved LCSR for the twist-two hadronic
photon correction to the $B \to \gamma$ form factors was subsequently constructed with the spectral representations
of the factorization formula (\ref{NLO factorization formula at LT}).
The subleading-power correction to the $B \to \gamma \ell \nu$ amplitude  from the leading-twist
photon DA was shown to preserve the symmetry relation between the two form factors due to the helicity conservation,
in agreement with the observation made in \cite{Ball:2003fq}.

Along the same vein, we proceed to compute the two-particle and three-particle higher-twist hadronic photon corrections
to the $B \to \gamma \ell \nu$ form factors at tree level, up to the twist-four accuracy. The symmetry relation between the
two form factors $F_V(n \cdot p)$ and $F_A(n \cdot p)$ was found to be violated by both the two-particle and  three-particle
higher-twist effects of the photon light-cone DAs.
In addition, our calculations explicitly indicate that the correspondence between the heavy-quark expansion and
the twist expansion is generally invalid for the soft contributions to the exclusive $B$-meson decays,
in analogy to the similar pattern observed in the context of the pion-photon form factor \cite{Agaev:2010aq}.

Adding up different pieces contributing to the $B \to \gamma \ell \nu$ amplitude, we further investigated the
phenomenological impacts of the subleading-power hadronic photon contributions,
employing the conformal expansion of the photon DAs at the ``P-wave" accuracy.
Numerically,  the NLL twist-two hadronic photon correction was estimated to give rise to
an approximately ${\cal O}(30 \%)$ reduction of the leading-power contribution,
computed from QCD factorization, with the default values of theory inputs.
By contrast, the higher-twist hadronic photon contributions at LO in ${\cal O}(\alpha_s)$
was found to be of minor importance at $3 \, {\rm GeV} \leq n \cdot p \leq m_B$,
albeit with the rather conservative uncertainty ranges for the  nonperturbative parameters
collected in Table \ref{tab of parameters for photon DAs}.
Moreover, we observed that the dominant radiative effect of the leading-twist hadronic photon
contribution comes from the NLO QCD correction instead of the QCD resummation  of the  parametrically
large logarithms $m_b^2/\Lambda^2$. To understand the systematic uncertainty from the truncation
of the Gegenbauer expansion at the second order, we explored the numerical impact of the fourth
moment of the leading-twist photon DA in a wide interval $\alpha_4(\mu_0) \in [-0.2, 0.2]$ and
observed that the dependence of the twist-two hadronic photon correction to the
$B \to \gamma \ell \nu$ form factors on $\alpha_4(\mu_0)$ was rather moderate at
$n \cdot p \geq 3 \, {\rm GeV}$, at least in the framework of the LCSR method.
Our main theory predictions for the $B \to \gamma \ell \nu$ form factors
with the uncertainties from  variations of different input parameters added in quadrature
 were displayed in figure  \ref{fig: uncertainties of B to gamma form factors} and the
 poor constraint on the first inverse moment of the $B$-meson DA $\lambda_B(\mu_0)$ brought about
one of the major uncertainties for the theory calculations. In this respect,  the improved
measurements of the partial branching fractions ${\cal BR}(B \to \gamma \ell \nu, \, E_{\gamma} \geq E_{\rm cut})$
with the tighter phase-space cut on the photon energy to validate the perturbative QCD calculations
from the Belle II experiment will be of value to provide solid constraints on the inverse moment  $\lambda_B(\mu_0)$,
when combined with the theory predictions including the power suppressed contributions of different origins.

Further improvements of the theory descriptions of the $B \to \gamma \ell \nu$ form factors in QCD
can be pursued in  distinct directions. First, it would be of interest to perform the NLO QCD corrections
to the twist-three hadronic photon corrections with the LCSR approach for a systematic understanding of
the higher-twist contributions. The technical challenge of accomplishing this task lies in the demonstration
of QCD factorization for the vacuum-to-$B$-meson correlation function (\ref{def: correlation function})
in the presence of the non-trivial mixing of the two-particle and three-particle light-ray operators
under the  QCD renormalization. Second, exploring the subleading-power contributions to the radiative
leptonic $B$-meson decay in the framework of SCET directly will be indispensable for deepening our understanding
of factorization properties for more complicated exclusive $B$-meson decays, where the rapidity divergences of the
convolution integrals entering the corresponding factorization formulae already emerge at leading power
in the heavy quark expansion.  Earlier attempts to address this ambitious question have been  undertaken in
different contexts (for an incomplete list, see for instance
\cite{Beneke:2003pa,Bell:2005gw,Beneke:2008pi,Chiu:2012ir}).
Third, computing the  subleading-power corrections to the $B \to \gamma$ form factors from
the higher-twist $B$-meson DAs will be of both conceptual and phenomenological value to investigate general properties
of the twist expansion in heavy-quark effective theory (see \cite{Wang:2016qii} for a preliminary discussion
with an incomplete decomposition of the three-particle vacuum-to-$B$-meson matrix element on the light-cone).
To this end, we will need to employ the RG  equations for these  higher-twist $B$-meson DAs
at one loop following the discussions presented in \cite{Braun:2017liq},
where the evolution equations of the twist-four  $B$-meson DAs at one loop were demonstrated to be completely
integrable and therefore can be solved exactly.
We  are therefore anticipating dramatic  progress
toward better understanding of the strong interaction dynamics of the radiative leptonic decay
$B \to \gamma \ell \nu$ in QCD.

\subsection*{Acknowledgements}

Y.M.W acknowledges support from the National Youth Thousand Talents Program,
the Youth Hundred Academic Leaders Program of Nankai University, and the NSFC with
Grant No. 11675082 and 11735010.
The work of Y.L.S is supported by Natural Science Foundation of Shandong Province,
China under Grant No. ZR2015AQ006.


\appendix


\section{Spectral representations}
\label{app:spectral resp}

Here we collect the dispersion representations of the various convolution integrals
entering the NLL factorization formula of the vacuum-to-photon correlation function
(\ref{NLL factorization formula for the twist-two correction}) for the sake of
constructing the LCSR for the $B \to \gamma \ell \nu$ form factors.

\begin{eqnarray}
&&  {1 \over \pi} \, {\rm Im}_{s} \, \int_0^1 \, dz \, \phi_{\gamma} (z, \mu) \,
{1 \over z \, r_3 +\bar z \, r_2 - 1 + i \, 0} \, {1-r_2 \over r_1 -r_2} \,
\ln \left ( {1-r_2 \over r_1 -r_2} \right ) \nonumber \\
&& =   \int_0^1 \, d z \, {\cal P}{1 \over r_3-r_2} \,
\ln \left ( {z \, r_3 + \bar z \, r_2 -1  \over 1-r_2} \right )  \,
\theta(z \, r_3 + \bar z \, r_2 -1) \, \phi_{\gamma}^{\prime}(z, \mu) \nonumber \\
&& \hspace{0.5 cm} + \, \int_0^1 \, d z \, {\cal P}{1 \over r_3-r_2} \, \theta(z \, r_3 + \bar z \, r_2 -1) \,
{ \phi_{\gamma} (z, \mu) \over z}  \,.   \\
\nonumber \\
&&   {1 \over \pi} \, {\rm Im}_{s} \, \int_0^1 \, dz \, \phi_{\gamma} (z, \mu) \,
{1 \over z \, r_3 +\bar z \, r_2 - 1 + i \, 0} \, {1-r_3 \over r_1 -r_3} \,
\ln \left ( {1-r_1 \over 1 -r_3} \right ) \nonumber \\
&& = \int_0^1 \, d z \, {\cal P} {1 \over r_3 -r_2} \,
\ln \bigg | {1- z \, r_3 - \bar z \, r_2 \over 1-r_3} \bigg | \,
\left [ \theta(z \, r_3 + \bar z \, r_2 -1) - \theta(r_3-1) \right ] \, \phi_{\gamma}^{\prime}(z, \mu)  \nonumber \\
&& \hspace{0.5 cm} -  \int_0^1 \, d z \, {\cal P} {1 \over r_3 -r_2} \,
\left [ \theta(z \, r_3 + \bar z \, r_2 -1) - \theta(r_3-1) \right ] \,
{\phi_{\gamma}(z, \mu) \over \bar z}  \,.  \\
\nonumber  \\
&&   {1 \over \pi} \, {\rm Im}_{s} \,\int_0^1 \, dz \, \phi_{\gamma} (z, \mu) \,
{1 \over z \, r_3 +\bar z \, r_2 - 1 + i \, 0} \, {1 \over 1-r_1}
= {\theta(r_3 -1) \over (r_3 -r_2)^2} \,\, \phi_{\gamma}^{\prime} \left ({1-r_2 \over r_3 -r_2}, \mu \right )  \,. \\
\nonumber  \\
&&   {1 \over \pi} \, {\rm Im}_{s} \,\int_0^1 \, dz \, \phi_{\gamma} (z, \mu) \,
{1 \over z \, r_3 +\bar z \, r_2 - 1 + i \, 0} \, {1- r_2 \over r_1 - r_2} \,
{\rm Li}_2 \left ( 1 - {1 - r_1 \over 1- r_2} \right ) \nonumber \\
&&  = {\theta(r_3 -1) \over r_2 -r_3} \, {\pi^2 \over 6} \, \phi_{\gamma} \left ({1-r_2 \over r_3 -r_2}, \mu \right )
+ \int_0^1 \, d z \, \phi_{\gamma}(z, \mu) \, \left [{\cal P} {1 \over z \, r_3 +\bar z \, r_2 - 1 }
- {\cal P} {1 \over z \, (r_3 - r_2) } \right ]   \nonumber \\
&& \hspace{0.3 cm} \times \, \ln \left (1 -  {1 - z \, r_3 - \bar z \, r_2 \over 1- r_2 } \right ) \,
\theta(z \, r_3 +\bar z \, r_2 - 1) \,. \\
\nonumber \\
&& {1 \over \pi} \, {\rm Im}_{s} \,\int_0^1 \, dz \, \phi_{\gamma} (z, \mu) \,
{1 \over z \, r_3 +\bar z \, r_2 - 1 + i \, 0} \, {1- r_3 \over r_1 - r_3} \,
{\rm Li}_2 \left ( 1 - {1 - r_1 \over 1- r_3} \right ) \nonumber \\
&&  = {\theta(r_3 -1) \over r_2 -r_3} \, {\pi^2 \over 6} \, \phi_{\gamma} \left ({1-r_2 \over r_3 -r_2}, \mu \right )
+ \int_0^1 \, d z \, \phi_{\gamma}(z, \mu) \, \left [{\cal P} {1 \over z \, r_3 +\bar z \, r_2 - 1 }
- {\cal P} {1 \over \bar z \, (r_2 - r_3) } \right ]   \nonumber \\
&& \hspace{0.3 cm} \times \, {\rm sgn}(r_2-r_3) \, \ln \left (1 -  {1 - z \, r_3 - \bar z \, r_2 \over 1- r_3 } \right ) \,
\theta \left ({ z \, r_3 +\bar z \, r_2 - 1 \over 1-r_3 } \right )   \,. \\
\nonumber \\
&& {1 \over \pi} \, {\rm Im}_{s} \,\int_0^1 \, dz \, \phi_{\gamma} (z, \mu) \,
{1 \over z \, r_3 +\bar z \, r_2 - 1 + i \, 0}  \,\,  {1-r_2 \over r_1 -r_2} \,\, \ln^2 (1-r_1) \nonumber \\
&& = \int_0^1 \, d z \, {\cal P}{1 \over r_3 -r_2} \, \left [ \ln^2 |1-z \, r_3 - \bar z \, r_2|
- {\pi^2 \over 3} \right ] \, \theta(z \, r_3 +\bar z \, r_2 - 1) \,  \phi_{\gamma}^{\prime}(z, \mu) \nonumber \\
&& \hspace{0.3 cm} + \int_0^1 \, d z \, { \phi_{\gamma}(z, \mu)  \over z} \, \bigg [ 2 \, {\cal P}{1 \over r_3 -r_2} \,
 \ln |1-z \, r_3 - \bar z \, r_2| \, \theta(z \, r_3 +\bar z \, r_2 - 1)  \nonumber \\
&& \hspace{0.3 cm}   + \, \delta(r_3 -r_2) \, \ln^2 (1-r_2) \bigg  ]  \,. \\
\nonumber \\
&& {1 \over \pi} \, {\rm Im}_{s} \,\int_0^1 \, dz \, \phi_{\gamma} (z, \mu) \,
{1 \over z \, r_3 +\bar z \, r_2 - 1 + i \, 0}  \,\,  {1-r_3 \over r_1 -r_3} \,\, \ln^2 (1-r_1) \nonumber \\
&& = \int_0^1 \, d z \, {\cal P}{1 \over r_3 -r_2} \, \left [ \ln^2 |1-z \, r_3 - \bar z \, r_2|
- {\pi^2 \over 3} \right ] \, \theta(z \, r_3 +\bar z \, r_2 - 1) \,  \phi_{\gamma}^{\prime}(z, \mu) \nonumber \\
&& \hspace{0.3 cm} - \int_0^1 \, d z \, { \phi_{\gamma}(z, \mu)  \over \bar z} \, \bigg [ 2 \, {\cal P}{1 \over r_3 -r_2} \,
 \ln |1-z \, r_3 - \bar z \, r_2| \, \theta(z \, r_3 +\bar z \, r_2 - 1)  \nonumber \\
&& \hspace{0.3 cm}   + \, \delta(r_3 -r_2) \, \ln^2 (1-r_2) \bigg  ]  \,. \\
\nonumber \\
&& {1 \over \pi} \, {\rm Im}_{s} \,\int_0^1 \, dz \, \phi_{\gamma} (z, \mu) \,
{1 \over z \, r_3 +\bar z \, r_2 - 1 + i \, 0}  \,\,  {1-r_2 \over r_1 -r_2} \, \ln^2 (1-r_2) \nonumber \\
&& = \left [ { \theta(r_3 -1)  \over r_2 -r_3} \,\phi_{\gamma} \left ({1-r_2 \over r_3-r_2}, \mu \right )
+ \delta(r_3 -r_2) \, \int_0^1 \, d z \, {\phi_{\gamma}(z, \mu) \over z}\, \right ]  \, \ln^2 (1-r_2)  \,. \\
\nonumber \\
&& {1 \over \pi} \, {\rm Im}_{s} \,\int_0^1 \, dz \, \phi_{\gamma} (z, \mu) \,
{1 \over z \, r_3 +\bar z \, r_2 - 1 + i \, 0}  \,\,  {1-r_3 \over r_1 -r_3} \, \ln^2 (1-r_3) \nonumber \\
&& = \int_0^1 \, d z \, \phi_{\gamma}(z, \mu) \, {r_2 \, (r_2 - 2) \, (1+z) + 2 \, z \over r_2 \, (1-r_2) \, z \, \bar z } \,
\left [ \delta(r_2 -r_3) \, \ln (1-r_2) - {\theta(r_3 -1)  \over r_2 -r_3} \,
\theta \left ( z - {1-r_2 \over r_3 -r_2} \right ) \right ] \, \nonumber \\
&& \hspace{0.3 cm}  + \, {6 \over r_3-r_2} \, \bigg \{ {\theta(r_3 -1) \over r_3 -r_2} \,
\phi_{\gamma}^{\prime} \left ( {1-r_2 \over r_3-r_2}, \mu \right )
+ {\theta(r_3 -1)  \over r_3 -r_2} \, \ln (r_3 -1) \, \phi_{\gamma}^{\prime} (z=1, \mu)  \nonumber \\
&& \hspace{0.3 cm}  - {1 \over r_3 -r_2} \, \int_0^1 \, d z \, \ln ( z \, r_3 +\bar z \, r_2 - 1 ) \,
\theta(z \, r_3 +\bar z \, r_2 - 1) \, \phi_{\gamma}^{\prime \prime} (z, \mu)  \, \bigg \}  \nonumber \\
&&  \hspace{0.3 cm}  + \,  {r_2  \over (1-r_2) (r_2 -r_3)} \, \int_0^1 \, d z \, \ln (z \, r_3 +\bar z \, r_2 - 1) \,
\theta(z \, r_3 +\bar z \, r_2 - 1) \, \phi_{\gamma}^{\prime} (z, \mu)   \nonumber \\
&&  \hspace{0.3 cm} - \, \int_0^1 dz \,\phi_{\gamma}(z, \mu) \,\,
{2 \, (z - 2 \, \bar z \, r_2) \over \bar z \, r_2 \, (z \, r_3 + \bar z \, r_2)} \,\,
 \theta(z \, r_3 +\bar z \, r_2 - 1) \,. \\
\nonumber  \\
&& {1 \over \pi} \, {\rm Im}_{s} \,\int_0^1 \, dz \, \phi_{\gamma} (z, \mu) \,
{1 \over z \, r_3 +\bar z \, r_2 - 1 + i \, 0}  \,
\left ( {2 -r_2 \over r_1 -r_2 } -{4 \over r_2} +2 \right )  \nonumber \\
&& = {\theta(r_3-1) \over r_2 -r_3} \, \phi_{\gamma} \left ( {1-r_2 \over r_3 -r_2} \,, \mu \right ) \,
\left [{2 -r_2 \over 4-r_2} - {4 \over r_2} + 2 \right ]
- \, \delta(r_3 -r_2)  \, \int_0^1 d z \,
{\phi_{\gamma}(z, \mu)  \over z} \,. \\
\nonumber  \\
&& {1 \over \pi} \, {\rm Im}_{s} \,\int_0^1 \, dz \, \phi_{\gamma} (z, \mu) \,
{1 \over z \, r_3 +\bar z \, r_2 - 1 + i \, 0}  \,
\left [ {2 \over r_3 (r_1 -r_3)} + {2 \, (r_1 -2) \over r_1 -r_3}
- {6 \over r_3} \right ] \, \ln(1-r_3) \nonumber \\
&& = \int_0^1 \, d z \,  \phi_{\gamma} (z, \mu) \,
\bigg \{ \left [ {2 (r_2 -1) \over \bar z \, r_2} \, \delta(r_3 -r_2)
+ {4 \, z \, \theta(r_3 -1) \over (1- \bar z \, r_2)(r_3 -r_2)} \,
\delta \left (z -{1-r_2 \over r_3 -r_2} \right ) \,  \right ] \, \ln |1-r_3|  \nonumber \\
&& \hspace{0.3 cm} - \, \theta(r_3 -1)  \, \left [ {2 \, (1-r_2) \over \bar z \, r_2} \, {1 \over r_3 -r_2}
- {2 \, (1 -3 \, \bar z \, r_2) \over  r_2 \, r_3 \, \bar z \,  (1- \bar z \, r_2)}
- {4 \, z \over 1 -\bar z \, r_2} \, {\cal P}{1 \over z \, r_3 + \bar z \, r_2 -1}\right ] \bigg \}   \,. \\
\nonumber  \\
&& {1 \over \pi} \, {\rm Im}_{s} \,\int_0^1 \, dz \, \phi_{\gamma} (z, \mu) \,
{1 \over z \, r_3 +\bar z \, r_2 - 1 + i \, 0}  \,
\left [ {r_2 \over 2 \, (r_1 -r_2)} - {3 \over 1-r_1} - {15 \over 2}\right ]  \nonumber \\
&& = {r_2  \over 2 \, (1-r_2)} \, \delta(r_2 -r_3)  \, \int_0^1 \, d z \, {\phi_{\gamma} (z, \mu) \over z}
- {3 \, \theta(r_3-1)\over (r_3-r_2)^2} \, \phi_{\gamma}^{\prime} \left ({1-r_2 \over r_3-r_2}, \mu \right )  \nonumber \\
&&  \hspace{0.3 cm} - \, {16 \, r_2 - 15 \over 2 \, (1-r_2) }  \, {\theta(r_3-1) \over r_3 -r_2} \,
\, \phi_{\gamma} \left ({1-r_2 \over r_3-r_2}, \mu \right )   \,.
\end{eqnarray}

\section{Higher-twist photon DAs}
\label{app:Higher-twist photon DAs}

In this Appendix we will collect the operator-level definitions of the two-particle and three-particle
photon  DAs on the light-cone up to the twist-four accuracy as presented in \cite{Ball:2002ps}.

\begin{eqnarray}
&& \langle \gamma(p) |\bar q(x) \, W_c(x, 0) \,\, \sigma_{\alpha \beta} \,\, q(0)| 0 \rangle  \nonumber \\
&& = i \, g_{\rm em} \, Q_q \,  \langle \bar q q \rangle(\mu) \,
(p_{\beta} \, \epsilon_{\alpha}^{\ast} - p_{\alpha} \, \epsilon_{\beta}^{\ast}) \,
\int_0^1 \, d z \, e^{i \, z \, p \cdot x} \, \left [  \chi(\mu) \, \phi_{\gamma}(z, \mu)
+ {x^2 \over 16} \, \mathbb{A}(z, \mu) \right ]   \nonumber \\
&& \hspace{0.3 cm} + \, {i \over 2} \, g_{\rm em} \, Q_q \,  {\langle \bar q q \rangle(\mu) \over q \cdot x} \,
(x_{\beta} \, \epsilon_{\alpha}^{\ast} - x_{\alpha} \, \epsilon_{\beta}^{\ast}) \,
\int_0^1 \, d z \, e^{i \, z \, p \cdot x} \, h_{\gamma}(z, \mu) \,. \\
\nonumber \\
&& \langle \gamma(p) |\bar q(x) \, W_c(x, 0) \,\, \gamma_{\alpha} \,\, q(0)| 0 \rangle
= - g_{\rm em} \, Q_q \, f_{3 \gamma}(\mu) \, \epsilon_{\alpha}^{\ast} \,
\int_0^1 \, d z \, e^{i \, z \, p \cdot x} \, \psi^{(v)}(z, \mu)  \,. \\
\nonumber \\
&& \langle \gamma(p) |\bar q(x) \, W_c(x, 0) \,\, \gamma_{\alpha} \, \gamma_5 \,\, q(0)| 0 \rangle \nonumber \\
&& = {g_{\rm em} \over 4} \, Q_q \,  f_{3 \gamma}(\mu) \, \varepsilon_{\alpha \beta \rho \tau} \,
p^{\rho} \, x^{\tau} \,\epsilon^{\ast \, \beta} \, \int_0^1 \, d z \, e^{i \, z \, p \cdot x} \,
\, \psi^{(a)}(z, \mu)\,. \\
\nonumber \\
&& \langle \gamma(p) |\bar q(x) \, W_c(x, 0) \,\, g_s \, G_{\alpha \beta}(v \, x) \, \, q(0)| 0 \rangle \nonumber \\
&& =  i \, g_{\rm em} \, Q_q \,  \langle \bar q q \rangle(\mu) \,
(p_{\beta} \, \epsilon_{\alpha}^{\ast} - p_{\alpha} \, \epsilon_{\beta}^{\ast}) \,
\int [{\cal D} \alpha_i] \, e^{i \, (\alpha_q +  \, v \, \alpha_g) \, p \cdot x} \, S(\alpha_i, \mu) \,. \\
\nonumber \\
&& \langle \gamma(p) |\bar q(x) \, W_c(x, 0) \,\, g_s \, \widetilde{G}_{\alpha \beta}(v \, x) \,
i \, \gamma_5 \,\,  q(0)| 0 \rangle \nonumber \\
&& =  i \, g_{\rm em} \, Q_q \,  \langle \bar q q \rangle(\mu) \,
(p_{\beta} \, \epsilon_{\alpha}^{\ast} - p_{\alpha} \, \epsilon_{\beta}^{\ast}) \,
\int [{\cal D} \alpha_i] \, e^{i \, (\alpha_q +  \, v \, \alpha_g) \, p \cdot x} \, \widetilde{S}(\alpha_i, \mu) \,. \\
\nonumber \\
&& \langle \gamma(p) |\bar q(x) \, W_c(x, 0) \,\, g_s \, \widetilde{G}_{\alpha \beta}(v \, x) \,
\gamma_{\rho} \, \gamma_5 \,\,  q(0)| 0 \rangle \nonumber \\
&& = - g_{\rm em} \, Q_q \, f_{3 \gamma}(\mu) \, p_{\rho} \,
(p_{\beta} \, \epsilon_{\alpha}^{\ast} - p_{\alpha} \, \epsilon_{\beta}^{\ast}) \,
\int [{\cal D} \alpha_i] \, e^{i \, (\alpha_q +  \, v \, \alpha_g) \, p \cdot x} \, A(\alpha_i, \mu) \,. \\
\nonumber \\
&& \langle \gamma(p) |\bar q(x) \, W_c(x, 0) \,\, g_s \, G_{\alpha \beta}(v \, x) \,
i \, \gamma_{\rho} \,\,  q(0)| 0 \rangle \nonumber \\
&& =  g_{\rm em} \, Q_q \, f_{3 \gamma}(\mu) \,  p_{\rho} \,
(p_{\beta} \, \epsilon_{\alpha}^{\ast} - p_{\alpha} \, \epsilon_{\beta}^{\ast}) \,
\int [{\cal D} \alpha_i] \, e^{i \, (\alpha_q +  \, v \, \alpha_g) \, p \cdot x} \, V(\alpha_i, \mu)  \,. \\
\nonumber \\
&& \langle \gamma(p) |\bar q(x) \, W_c(x, 0) \,\, g_{\rm em} \, Q_q \,  F_{\alpha \beta}(v \, x) \, \, q(0)| 0 \rangle \nonumber \\
&& =  i \, g_{\rm em} \, Q_q \,  \langle \bar q q \rangle(\mu) \,
(p_{\beta} \, \epsilon_{\alpha}^{\ast} - p_{\alpha} \, \epsilon_{\beta}^{\ast}) \,
\int [{\cal D} \alpha_i] \, e^{i \, (\alpha_q +  \, v \, \alpha_g) \, p \cdot x} \, S_{\gamma}(\alpha_i, \mu) \,.  \\
\nonumber \\
&& \langle \gamma(p) |\bar q(x) \, W_c(x, 0) \,\, \sigma_{\rho \tau} \,\, g_s \, G_{\alpha \beta}(v \, x)
\,\,  q(0)| 0 \rangle \nonumber \\
&& = - \, g_{\rm em} \, Q_q \,\langle \bar q q \rangle(\mu) \,
\left [p_{\rho} \, \epsilon_{\alpha}^{\ast} \, g_{\tau \beta}^{\perp}
- p_{\tau} \, \epsilon_{\alpha}^{\ast} \, g_{\rho \beta}^{\perp} - (\alpha \leftrightarrow \beta) \right ]  \,
\int [{\cal D} \alpha_i] \, e^{i \, (\alpha_q +  \, v \, \alpha_g) \, p \cdot x} \, T_{1}(\alpha_i, \mu)  \nonumber \\
&& \hspace{0.4 cm} - \, g_{\rm em} \, Q_q \,\langle \bar q q \rangle(\mu) \,
\left [p_{\alpha} \, \epsilon_{\rho}^{\ast} \, g_{\tau \beta}^{\perp}
- p_{\beta} \, \epsilon_{\rho}^{\ast} \, g_{\tau \alpha}^{\perp} - (\rho \leftrightarrow \tau) \right ]  \,
\int [{\cal D} \alpha_i] \, e^{i \, (\alpha_q +  \, v \, \alpha_g) \, p \cdot x} \, T_{2}(\alpha_i, \mu) \nonumber \\
&& \hspace{0.4 cm} - \, g_{\rm em} \, Q_q \,\langle \bar q q \rangle(\mu) \,
\frac{(p_{\alpha} \, x_{\beta} - p_{\beta} \, x_{\alpha} ) (p_{\rho} \, \epsilon_{\tau}^{\ast} - p_{\tau} \,\epsilon_{\rho}^{\ast})}
{p \cdot x} \, \int [{\cal D} \alpha_i] \, e^{i \, (\alpha_q +  \, v \, \alpha_g) \, p \cdot x} \,
 T_{3}(\alpha_i, \mu) \nonumber \\
&& \hspace{0.4 cm} - \, g_{\rm em} \, Q_q \,\langle \bar q q \rangle(\mu) \,
\frac{(p_{\rho} \, x_{\tau} - p_{\tau} \, x_{\rho} ) (p_{\alpha} \, \epsilon_{\beta}^{\ast} - p_{\beta} \,\epsilon_{\alpha}^{\ast})}
{p \cdot x} \, \int [{\cal D} \alpha_i] \, e^{i \, (\alpha_q +  \, v \, \alpha_g) \, p \cdot x} \,
 T_{4}(\alpha_i, \mu) \,. \\
 \nonumber \\
&& \langle \gamma(p) |\bar q(x) \, W_c(x, 0) \,\, \sigma_{\rho \tau} \,\, g_{\rm em} \, Q_q \, F_{\alpha \beta}(v \, x)
\,\,  q(0)| 0 \rangle \nonumber \\
&& = - \, g_{\rm em} \, Q_q \,\langle \bar q q \rangle(\mu) \,
\frac{(p_{\rho} \, x_{\tau} - p_{\tau} \, x_{\rho} ) (p_{\alpha} \, \epsilon_{\beta}^{\ast} - p_{\beta} \,\epsilon_{\alpha}^{\ast})}
{p \cdot x} \, \int [{\cal D} \alpha_i] \, e^{i \, (\alpha_q +  \, v \, \alpha_g) \, p \cdot x} \,
 T_{4}^{\gamma}(\alpha_i, \mu)  + ... \,
\end{eqnarray}
Here,  we have employed the following notations for the dual field strength tensor
and the integration measure
\begin{eqnarray}
\widetilde{G}_{\alpha \beta}= {1 \over 2} \,  \varepsilon_{\alpha \beta \rho \tau }  \, G^{\rho \tau} \,,
\qquad \int [{\cal D} \alpha_i] \equiv \int_0^1 d \alpha_q \, \int_0^1 d \alpha_{\bar q} \,
\int_0^1 d \alpha_g \, \delta \left (1-\alpha_q - \alpha_{\bar q} -\alpha_g \right )\,.
\end{eqnarray}


\end{document}